\documentclass[preprint,authoryear,11pt,review]{elsarticle}
\usepackage{setspace}
\usepackage{lscape}
\usepackage[nodots]{numcompress}
\usepackage{amssymb,epsfig}
\usepackage{gensymb}
\usepackage{fancyhdr}
\usepackage{color,colordvi}
\usepackage{amsmath}
\usepackage{mathrsfs}
\usepackage{placeins}
\usepackage{accents}
\usepackage{float}
\usepackage{subfig}
\usepackage{epstopdf}
\usepackage[table]{xcolor}

\usepackage{rotating}
\usepackage{multirow}
\usepackage{graphicx} 
\usepackage{bm}

\begin{document}
\def\finp{\\ \hspace*{0.5cm}}
\def\r {{\bf r}}
\def\sc {\hat{\sigma}}
\def\tT {\delta T}
\def \tD {\delta D}
\def \tF {\delta F}
\def \tx {\delta x}
\def\tsig {\delta \sigma}
\def \ps {\delta s}
\def \tp {\delta p}
\def\eps {\epsilon^{'}}
\def\bsigma {\boldsymbol{\sigma}}
\def\btd{\boldsymbol{\dot{T}}}
\def\bt{\boldsymbol{T}}
\def\bl{\boldsymbol{L}}
\def\msr {  \underline{\underline{D}}}
\def\ml {  \underline{\underline{L}}}
\def\mdl {  \underline{\underline{\dot{L}}}}
\def\mo{  \underline{\underline{\Omega}}}
\def\mdsr {  \underline{\underline{\dot{D}}}}
\def\sr { \underline{\underline{d}}}
\def\ts {\underline{\underline{\sigma}}}
\def\tsm {\underline{\underline{\Sigma}}}
\def\smij {\Sigma_{ij}}
\def\smik {\Sigma_{ik}}
\def\sij {\sigma_{ij}}
\def\avsij {< \sij >}
\def\mc{\underline{\underline{C}}}
\def\sik {\sigma_{ik}}
\def\uvol {\frac{1}{\vert \Omega \vert}}
\def\dij {d_{ij}}
\def\do {\partial \Omega}
\def\into {\int_{\Omega}}
\def\intdo {\int_{\partial \Omega}}
\def\dmij {D_{ij}}
\def\dmik {D_{ik}}
\def\ux {\underline{x}}
\def\uv {\underline{v}}
\def\work {< \ts : \delta \sr >}
\def\works { < \ts : \sr >}
\def\bibsection{\section*{References}}
\doublespacing

\begin{frontmatter}

\journal{Mechanics of Materials}

\title{The effect of initial texture on multiple necking formation in polycrystalline thin rings subjected to dynamic expansion}

\author[UC3M]{K. E. N'souglo}
\author[IPPT]{K. Kowalczyk-Gajewska}
\author[HZH]{M. Marvi-Mashhadi}
\author[UC3M]{J. A. Rodr{\'i}guez-Mart{\'i}nez \corref{cor}}
\cortext[cor]{Corresponding author. E-mail address: jarmarti@ing.uc3m.es}

\address[UC3M]{Department of Continuum Mechanics and Structural Analysis. University Carlos III of Madrid. Avda. de la Universidad, 30. 28911 Legan{\'e}s, Madrid, Spain}
\address[IPPT]{Institute of Fundamental Technological Research. Polish Academy of Sciences. Pawińskiego 5b, 02 106 Warsaw, Poland}
\address[HZH]{Institute of Material Systems Modeling. Helmholtz-Zentrum Hereon. Max-Planck Straße 1, 21502, Geesthacht, Germany}

\begin{abstract}

In this paper, we have investigated, using finite element calculations, the effect of initial texture on the formation of multiple necking patterns in ductile metallic rings subjected to rapid radial expansion. The mechanical behavior of the material has been modeled with the elasto-viscoplastic single crystal constitutive model developed by \citet{marin2006}. The polycrystalline microstructure of the ring has been generated using random Voronoi seeds. Both $5000$ grain and $15000$ grain aggregates have been investigated, and for each polycrystalline aggregate three different spatial distributions of grains have been considered. The calculations have been performed within a wide range of strain rates varying from $1.66 \cdot 10^4 ~ \text{s}^{-1}$ to $3.33 \cdot 10^5 ~ \text{s}^{-1}$, and the rings have been modeled with four different initial textures: isotropic texture, $\left\langle 001\right\rangle\parallel\Theta$ Goss texture, $\left\langle 001\right\rangle\parallel$ R Goss texture and $\left\langle 111\right\rangle\parallel$ Z fiber texture. The finite element results show that: (i) the spatial distribution of grains affects the location of the necks, (ii) the decrease of the grain size delays the formation of the necking pattern and increases the number of necks, (iii) the initial texture affects the number of necks, the location of the necks, and the necking time, (iv) the development of the necks is accompanied by a local increase of the slip activity. This work provides new insights into the effect of crystallographic microstructure on dynamic plastic localization and guidelines to tailor the initial texture in order to delay dynamic necking formation and, thus, to improve the energy absorption capacity of ductile metallic materials at high strain rates.

\end{abstract}


\begin{keyword}

Dynamic necking \sep Inertia \sep Crystal plasticity \sep Texture \sep Finite elements

\end{keyword}

\end{frontmatter}

\section{Introduction}
\label{Introduction}

The ring expansion experiment developed by \cite{Niordson65} has become a reference benchmark problem to investigate dynamic necking localization and fragmentation of ductile metallic materials. The test consists of a thin circular specimen with square/rectangular cross-section which is expanded radially at high strain rates using electromagnetic loading schemes, explosive charges or mechanical loading systems such as gas guns \citep{Fyfe80,Goto08,WOOD2021109908,gant2021plate}, so that multiple necks are formed throughout the circumference of the sample at large strains, leading to the final fragmentation of the ring. The principal advantage of the rapidly expanding ring experiment --in comparison with the tensile impact testing of linear specimens (slender bars)-- is that the radial symmetry of both specimen and loading nearly eliminates the propagation of stress waves along the hoop direction of the sample before necking localization, thus revealing the true dynamic properties of the material.

\

The experiments of \cite{Grady83} with OFHC copper and aluminum $1100$-O rings having inner diameter of $32~\text{mm}$ and square cross-section of $1~\text{mm}^2$, tested at radial velocities ranging from about $20~\text{m}/\text{s}$ to $200~\text{m}/\text{s}$, showed that the number of necks, the proportion of necks developing into fractures, and the fracture strain, increase with the applied velocity. Following \cite{Mott47} postulates, the distribution of neck spacings and fragment sizes was attributed to the spatial variability of the necking and fracture strains throughout the circumference of the ring, and to the unloading waves released from growing necks and early fractures which remove the driving force from nearby sections of the specimen before fracture is completed. Moreover, \cite{Altynova96} carried out ring expansion tests with aluminum $6061$ and OFHC copper samples, for radial speeds ranging between $50~\text{m}/\text{s}$ and $300~\text{m}/\text{s}$. The rings had inner diameter of $30.5~\text{mm}$, radial thickness of $1~\text{mm}$, and different widths varying from $1~\text{mm}$ to $2~\text{mm}$. For expansion velocities above $200~\text{m}/\text{s}$, the ductility of aluminum 6061 and OFHC copper rings increased with respect to the quasi-static value by $60\%$ and $250\%$, respectively. The improved ductility at high strain rates was attributed to inertia effects which diffuse deformation throughout the specimen, leading to stabilization against neck growth and delaying failure. Ring expansion experiments on U6N specimens were carried out by \cite{Grady03}. The samples had inner diameter of $30~\text{mm}$ and square cross-section of $0.75 \times 0.75 ~\text{mm}^2$, and were submitted to radial velocities ranging from $50~\text{m}/\text{s}$ to $300~\text{m}/\text{s}$. The number of fragments, and the distributions of fragment sizes and masses, were compared with the fragmentation theories of \cite{Mott47} and \cite{Kipp85} which consider the fragmentation patterns of metallic samples subjected to intense impulsive tensile loads to be the result of the activation of weak points of the material (e.g., material defects like pores, cavities and cracks formed during materials processing, and microstructural heterogeneities like soft phases, grain boundaries) which are scattered throughout the specimen and determine the distributions of necks and fracture sites. \cite{Zhang06} tested aluminum 6061-O specimens with inner diameter of $30.5~\text{mm}$, and rectangular cross-section of thickness $0.5~\text{mm}$ and length $1~\text{mm}$. The radial loading speeds investigated ranged between $80~\text{m}/\text{s}$ and $200~\text{m}/\text{s}$. Consistent with previous works of \cite{Grady83} and \cite{Grady03}, the experiments of \cite{Zhang06} showed that the neck and fragment distributions shift towards smaller neck spacings and smaller fragments, and towards narrower distributions, with increasing loading rate. It was concluded that as the strain rate increases, the propagation of the release wave emanating from growing necks and early fractures is limited to shorter times, so that the unloading wave does not travel far enough quickly to inhibit further nucleation of necks and fractures at neighboring locations. An incomplete list of additional papers including ring expansion experiments on metallic samples is \citep{perrone1968use,stepanov1984experimental,Janiszewski12,CLICHE2018189}.

\

The ring expansion problem has been also extensively investigated over the last two decades using finite element calculations. For instance, \cite{Guduru02} carried out simulations in ABAQUS/Explicit of the ring expansion experiments performed by \cite{Grady83}, modeling the mechanical behavior of the material using Gurson porous plasticity, and assuming that fracture occurs when a critical value of porosity is achieved. The finite element model consisted of a long rod submitted to rapid stretching and with initial conditions consistent with the expanding ring. The finite element results yielded qualitative agreement with the experiments for the number of necks, the fragmentation statistics, and the increase in the fracture strain with the loading rate. Moreover, \cite{Rusinek07} performed ring expansion finite element simulations modeling the material behavior with von Mises plasticity, different constitutive relations for the evolution of the yield stress, and a fracture criterion which assumed the material to fail when a predefined value of plastic strain is reached. The numerical predictions for the number of fragments were compared with the experimental data reported by \cite{Grady83}, \cite{Altynova96} and \cite{diep2004fragmentation} for aluminum 1100-0, aluminum 6061 and steel 34CrNiMo6, respectively, and satisfactory quantitative agreement was obtained for the whole range of loading velocities considered. Moreover, additional calculations with baseline material parameters corresponding to mild steel showed that the number of fragments increases with the decrease of the strain hardening and with the increase of the material density. Shortly after, \cite{Zhang10a} simulated ring expansion experiments on aluminium 6061-O specimens with different cross-section sizes using von Mises plasticity and a power law strain hardening to model the mechanical behavior of the material. A random distribution of material defects (modeled as elements of the finite element grid with low yield stress) was included in the finite element model to break the symmetry of the problem and trigger plastic localization. The loading condition was a pulse pressure applied in the inner surface of the ring, leading to a maximum expansion velocity of $250~\text{m}/\text{s}$, corresponding to {a} strain rate of $10^4~\text{s}^{-1}$. The calculations showed, in agreement with the experiments, that increasing the cross-section of the ring delays localized necking. \cite{nsouglo2018} carried out finite element calculations of elasto-plastic bars with yielding modeled with Gurson plasticity and submitted to dynamic tension at strain rates ranging from $10^3~\text{s}^{-1}$ to $5\cdot10^4~\text{s}^{-1}$. An initial velocity field compatible with the loading condition was applied on the specimen to minimize the intervention of waves before plastic flow localization, so that the deformation field during the bar extension was consistent with the ring expansion problem. The main difference with respect to the calculations of \cite{Guduru02} is that the initial porosity --which in the Gurson model intends to idealize material defects-- was randomly varied at the element level, using predefined upper and lower bounds for the porosity variation, and different spatial distributions. The results of \cite{nsouglo2018} showed that, for the lower strain rates, the distribution of neck spacings is heterogeneous, and the porosity defects act as preferential sites for the nucleation of necks. On the other hand, as the strain rate increases, the distribution of neck spacings becomes more homogeneous due to the regularizing effect of inertia. Moreover, \cite{vaz2019comparative} performed a comparative study on the formation of multiple necks and fragments in elasto-plastic and hyperelastic rings subjected to expansion velocities ranging between $25~\text{m}/\text{s}$ and $600~\text{m}/\text{s}$. The elasto-plastic material and the hyperelastic material were modeled with constitutive equations which provide nearly the same stress-strain response during monotonic uniaxial tensile loading, and fracture was assumed to occur at the same level of deformation energy. The calculations predicted virtually the same number of necks for the elasto-plastic and the hyperelastic rings, however, the mechanisms controlling the development of the necking pattern and the final fragmentation of the specimens were shown to be different. In the elasto-plastic rings several necks were arrested due to the release waves that travel throughout the circumference of the specimen after the localization process has started, so that the number of fractures was significantly lower than the number of necks. On the other hand, the release waves did not arrest the growth of any neck in the hyperelastic rings, on the contrary, the elastic energy released from the sections of the ring which are unloading during the localization process fueled the development of the necks, so that most of the necks developed into fractures. More recently, \cite{MARVIMASHHADI2021102999} performed finite element calculations of expanding rings which included explicit representation of the porous microstructure of different additive-manufactured metals. The mechanical behavior of the material was modeled with von Mises plasticity, and a critical value of effective plastic strain was used as fracture criterion. The calculations, carried out for expansion velocities ranging from $50~\text{m}/\text{s}$ to $500~\text{m}/\text{s}$, yielded individualized correlations between the number of necks and fragments, and the main features of the porous microstructure, including the initial void volume fraction and the maximum void size. The following is an incomplete list of additional finite element works in which the formation of multiple necks and fragments in metallic rings subjected to dynamic expansion is modeled \citep{pandolfi1999finite,becker2002ring,Rodriguez12c,Vaz17,MARVIMASHHADI2020103661,ELMAI2022104798}. 

\

Notice that most of the works cited in previous paragraph focused the attention on the effect that mechanical behavior of the material and material defects have on the necking and fragmentation patterns. However, all these papers modeled the material response using macroscopic constitutive equations, paying no attention to the polycrystalline microstructure of the metallic rings, although the microstructure is known to be responsible for inherent material variations at the micromechanical level that favor plastic localization. This is precisely the gap we intend to fill in this paper, in which crystal plasticity finite element simulations of metallic rings subjected {to} dynamic expansion have been performed using the elasto-viscoplastic constitutive model developed by \citet{marin2006}. The polycrystalline microstructure of the ring has been generated using two different random Voronoi tessellations containing $5000$ and $15000$ grains, respectively. The grains of the microstructure have been assigned with an initial crystal orientation corresponding to one of the four different textures investigated: isotropic texture, $\left\langle 001\right\rangle\parallel\Theta$ Goss texture, $\left\langle 001\right\rangle\parallel$ R Goss texture and $\left\langle 111\right\rangle\parallel$ Z fiber texture. A salient feature of this investigation is that the finite element results show that the initial texture can be tailored to delay the formation of dynamic necks, improving the energy absorption capacity of ductile metallic materials at high strain rates. Note that, while \cite{dequiedt2021localization} were probably the first researchers to perform crystal plasticity finite element simulations in order to study the effect of material microstructure on dynamic plastic localization, to the authors' knowledge, this is the first paper ever providing a systematic analysis on the effect of initial texture on the formation of multiple necking patterns.

\section{Crystal plasticity model}
\label{Crystal plasticity constitutive framework}

The single crystal constitutive model used to perform the calculations of Section \ref{Results} is the elasto-viscoplastic formulation developed by \citet{marin2006}, specialized to the case of small elastic strains. The main features of the formulation are presented in what follows, while the reader is referred to the technical report of \citet{marin2006} {to obtain} additional details. The model formulation follows the classical works by \citet{Hill72} and \citet{Asaro85} as concerns the kinematics description and the use of the rate-dependent power law for slip.

\

Accordingly, the constitutive model considers that crystallographic slip is the dominant deformation mechanism. The single crystal kinematics {rely} on the multiplicative decomposition of the deformation gradient $\mathbf{F}$ into an elastic component $\mathbf{F}^{e}$ and a plastic component $\mathbf{F}^{p}$:

\begin{eqnarray}
	\mathbf{F}=\mathbf{F}^{e}\mathbf{F}^{p}=\mathbf{V}^{e}\mathbf{R}^{e}\mathbf{F}^{p}=\mathbf{V}^{e}\hat{\mathbf{F}}^{p}
	\label{Decomposition_of_F}
\end{eqnarray}

\noindent where $\mathbf{F}^{p}$ describes the motion of dislocations on crystallographic planes (leaving the crystal lattice unchanged), while $\mathbf{R}^{e}$ and $\mathbf{V}^{e}$ represent the rotation and the elastic stretching of the lattice, respectively. The decomposition \eqref{Decomposition_of_F} introduces two intermediate configurations between the reference $\mathcal{B}_{0}$ and the current $\mathcal{B}$ configurations, which are denoted as $\bar{\mathcal{B}}$ and $\tilde{\mathcal{B}}$, respectively. Note that $\mathbf{F}^{p}$ brings the crystal from the reference $\mathcal{B}_{0}$ to the intermediate configuration $\bar{\mathcal{B}}$, while $\hat{\mathbf{F}}^{p}$ connects $\mathcal{B}_{0}$ with the intermediate configuration $\tilde{\mathcal{B}}$, which is used to express the crystal constitutive equations.

\

Moreover, assuming small elastic strains, $\mathbf{V}^{e}$ is approximated by:

\begin{eqnarray}
	\mathbf{V}^{e}=\mathbf{1}+\bm{\epsilon}^{e}
	\label{Small_elastic_strain_assumption}
\end{eqnarray}

\noindent where $\mathbf{1}$ is the unit second-order tensor and $\bm{\epsilon}^{e}$ is the small elastic strain tensor with $\left\| \bm{\epsilon}^{e} \right\| \ll 1$, so that the current configuration $\mathcal{B}$ differs from the intermediate configuration $\tilde{\mathcal{B}}$ only by an infinitesimal amount. Hence, the rate of deformation tensor $\mathbf{d}$ and the spin tensor $\mathbf{w}$ are expressed as:

\begin{eqnarray}
	\begin{split}
			& \mathbf{d}=\accentset{\triangledown}{\bm{\epsilon}}^{e}+\tilde{\mathbf{D}}^{p}\\
			& \mathbf{w}=-\mathrm{skew}\left(\dot{\bm{\epsilon}}^{e}\bm{\epsilon}^{e}\right)+\tilde{\mathbf{\Omega}}^{e}+\tilde{\mathbf{W}}^{p}
		\end{split}
	\label{}
\end{eqnarray}

\noindent with $\accentset{\triangledown}{\bm{\epsilon}}^{e}=\dot{\bm{\epsilon}}^{e}+\bm{\epsilon}^{e}\tilde{\mathbf{\Omega}}^{e}-\tilde{\mathbf{\Omega}}^{e}\bm{\epsilon}^{e}$ and $\tilde{\mathbf{\Omega}}^{e}=\dot{\mathbf{R}}^{e}\left( \mathbf{R}^{e}\right) ^T$ being the elastic part of the rate of deformation tensor and the elastic lattice spin tensor, respectively. Note that the superscript $\left(\right)^T$ refers to the tensor transpose, $\accentset{\triangledown}{\left( \right) }$ is the material Jaumann rate, and $\dot{\left(\right)}$ denotes differentiation with respect to time. Moreover, $\tilde{\mathbf{D}}^{p}$ and $\tilde{\mathbf{W}}^{p}$ are the plastic part of the rate of deformation tensor and the plastic spin tensor given by:

\begin{eqnarray}
	\begin{split}
		& \tilde{\mathbf{D}}^{p}=\sum_{\alpha=1}^{N}\dot{\gamma}^{\alpha}\mathrm{sym}(\tilde{\mathbf{Z}}^{\alpha})\\
		& \tilde{\mathbf{W}}^{p}=\sum_{\alpha=1}^{N}\dot{\gamma}^{\alpha}\mathrm{skew}(\tilde{\mathbf{Z}}^{\alpha})
	\end{split}
	\label{}
\end{eqnarray}

\noindent where $\dot{\gamma}^{\alpha}$ is the plastic shear rate on the $\alpha$ slip system, $N$ is the number of slip systems, and $\tilde{\mathbf{Z}}^{\alpha}=\tilde{\mathbf{s}}^{\alpha}\otimes\tilde{\mathbf{m}}^{\alpha}$ is the Schmidt tensor with $\tilde{\mathbf{s}}^{\alpha}=\mathbf{R}^{e}\mathbf{s}^{\alpha}$ and $\tilde{\mathbf{m}}^{\alpha}=\mathbf{R}^{e}\mathbf{m}^{\alpha}$ being the slip direction and the {normal to the} slip plane in the current configuration, respectively. 

\

The plastic shear rate is defined with a power law flow rule:

\begin{eqnarray}
	\dot{\gamma}^{\alpha}=\dot{\gamma}_{0}\left(\dfrac{\left|\tau^{\alpha}\right|}{\kappa^{\alpha}}\right)^{\frac{1}{m}}\mathrm{sign}\left(\tau^{\alpha}\right) 
	\label{Flow_rule}
\end{eqnarray}

\noindent where $\dot{\gamma}_{0}$ is a reference shear strain rate and $m$ is the rate sensitivity of slip. Moreover, $\kappa^{\alpha}$ is the $\alpha$ slip system hardness governed by the following modified Voce-type relation:

\begin{eqnarray}
	\dot{\kappa}^{\alpha}=h_{0}\left(\dfrac{\kappa_{s}-\kappa^{\alpha}}{\kappa_{s}-\kappa_{0}}\right)\dot{\gamma}
	\label{Hardening_law}
\end{eqnarray}

\noindent where $h_{0}$ is the initial hardening rate and $\kappa_{0}$ is the initial slip strength, while $\kappa_{s}$, the saturation slip strength, is given by:

\begin{eqnarray}
\kappa_{s}=\kappa_{s0}\left(\dfrac{\dot{\gamma}}{\dot{\gamma}_{s0}}\right)^{m'}
	\label{Hardening_law_2}
\end{eqnarray}

\noindent with $\dot{\gamma}=\sum_{\alpha=1}^{N}\left|\dot{\gamma}^{\alpha}\right|$ being the net shear strain rate within the crystal, and 
$\kappa_{s0}$, $\dot{\gamma}_{s0}$ and $m'$ being the slip system hardening parameters which, similarly to $h_{0}$ and $\kappa_{0}$, are taken to be the same for all slip systems.

\

Moreover, in equation \eqref{Flow_rule}, $\tau^{\alpha}$ is the resolved shear stress on the $\alpha$ slip system:

\begin{eqnarray}
	\tau^{\alpha}=\bm{\tau}:\mathrm{sym}(\tilde{\mathbf{Z}}^{\alpha})
	\label{}
\end{eqnarray}

\noindent where $\bm{\tau}$ is the Kirchhoff stress tensor defined by the elasticity relationship:

\begin{eqnarray}
	\bm{\tau}=\tilde{\mathbb{C}}^{e}:\bm{\epsilon}^{e}
	\label{Elasticity_relationship}
\end{eqnarray}

\noindent with $\tilde{\mathbb{C}}^{e}$ being the fourth order crystal elasticity tensor.

%

\

The calculations presented in Section \ref{Results} have been performed with the UMAT user subroutine developed by \citet{marin2006} for ABAQUS/Standard. The integration of the crystal plasticity constitutive model has been carried out using an implicit numerical integration procedure, see \citet{marin2006} for details. The material investigated is aluminum with FCC structure. The twelve $\left\lbrace1 1 1\right\rbrace\left\langle1 1 0\right\rangle$ slip systems in FCC crystal structure are shown in Table \ref{Slip_systems}. The anisotropic elasticity constants (see Eq. \eqref{Elasticity_relationship}), taken from \cite{marin2006}, and the parameters for the flow rule (see Eq. \eqref{Flow_rule}) and the hardening law (see Eq. \eqref{Hardening_law}), taken from \cite{beaudoin1994}, are shown in Table \ref{Material_parameters}. The initial density used in the calculations is $\rho=2710~\mathrm{kg/m}^{3}$. {Note that the mild strain rate sensitivity of the material presumably heightens the influence of texture on the multiple necking patterns. It is expected that increasing the value of the strain rate sensitivity parameter will decrease the relative influence of texture on necking ductility (strain rate sensitivity stabilizes the material behavior, see \cite{Mercier03} and \cite{Zhou06}). Nevertheless, validation of this hypothesis is left for future work.}

\

\begin{table}[hbtp]
	\begin{center}
			\begin{tabular}{c c c}
					\hline
					\rowcolor{gray!15} $\alpha$ & $\mathbf{m}^{\alpha}$ & $\mathbf{s}^{\alpha}$ \\ 
					\hline
					\hline
					 $1$ & $(1  1  1)$ & $[0  1 \bar{1}]$ \\
					 $2$ & $(1  1  1)$ & $[1  0 \bar{1}]$ \\
					 $3$ & $(1  1  1)$ & $[1 \bar{1}  0]$ \\
					 $4$ & $(\bar{1}  1  1)$ & $[0  1 \bar{1}]$ \\
					 $5$ & $(\bar{1}  1  1)$ & $[1  0  1]$ \\
					 $6$ & $(\bar{1}  1  1)$ & $[1  1  0]$ \\
					 $7$ & $(\bar{1} \bar{1}  1)$ & $[0  1  1]$ \\
					 $8$ & $(\bar{1} \bar{1}  1)$ & $[1  0  1]$ \\
					 $8$ & $(\bar{1} \bar{1}  1)$ & $[1 \bar{1}  0]$ \\
					 $10$ & $(1 \bar{1}  1)$ & $[0  1  1]$ \\
					 $11$ & $(1 \bar{1}  1)$ & $[1  0 \bar{1}]$ \\
					 $12$ & $(1 \bar{1}  1)$ & $[1  1  0]$ \\
					\hline
				\end{tabular}
		\end{center}
	\begin{center}
			\caption{Slip systems $\left(\mathbf{m}^{\alpha},\mathbf{s}^{\alpha}\right)$ in FCC crystal structure.}
			\label{Slip_systems}
		\end{center}
\end{table}

\begin{table}[hbtp]
	\begin{center}
			\begin{tabular}{c c c c c}
					\hline
					\rowcolor{gray!15} \textbf{Symbol} & \textbf{Value} \\
					\hline
					\hline
					$C_{11}$ & $108.2~\mathrm{GPa}$  \\
					$C_{12}$ & $61.3~\mathrm{GPa}$ \\
					$C_{44}$ & $28.5~\mathrm{GPa}$  \\
					\hline
					$\dot{\gamma}_{0}$ & $1~\mathrm{s}^{-1}$ \\
					$m$ & $0.05$ \\
					\hline
					$h_{0}$ & $200~\mathrm{MPa}$ \\
					$\kappa_{0}$ & $210~\mathrm{MPa}$ \\
					$\kappa_{s0}$ & $330~\mathrm{MPa}$ \\
					$\dot{\gamma}_{s0}$ & $5\times10^{10}~\mathrm{s}^{-1}$ \\
					$m'$ & $0.005$ \\
					\hline
				\end{tabular}
		\end{center}
	\begin{center}
			\caption{Material parameters for aluminum. {The anisotropic elasticity constants are taken from \cite{marin2006}, and the parameters for the flow rule and the hardening law from \cite{beaudoin1994}.}}
			\label{Material_parameters}
		\end{center}
\end{table}

\section{Finite element model}
\label{Finite element model}

\begin{figure}[hbtp]
\centering
\subfloat[]{	
	\includegraphics[width=17cm]{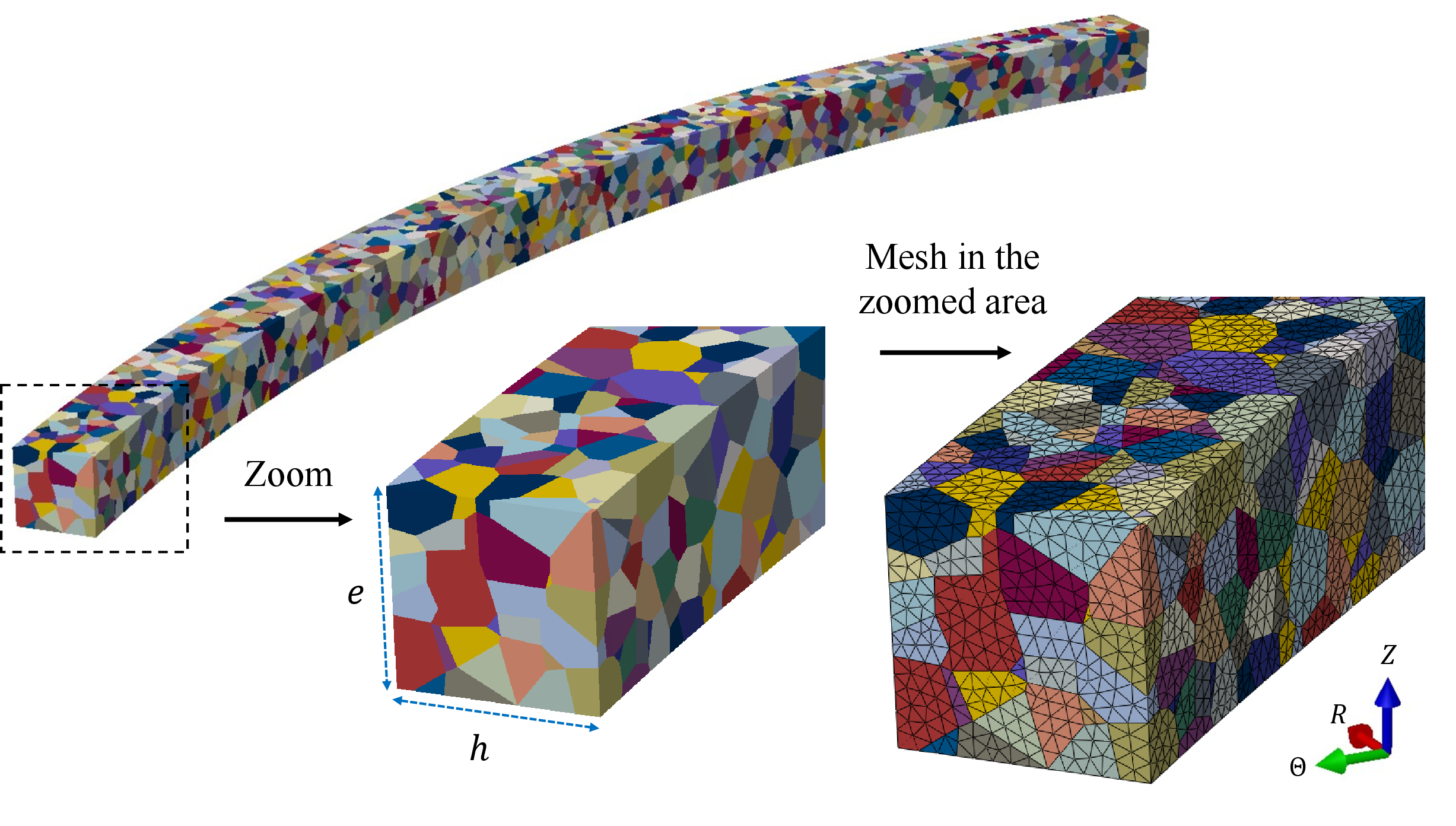}
	\label{POLYCRYSTAL_FEM}
}\\	
\subfloat[]{	
	\includegraphics[width=14cm]{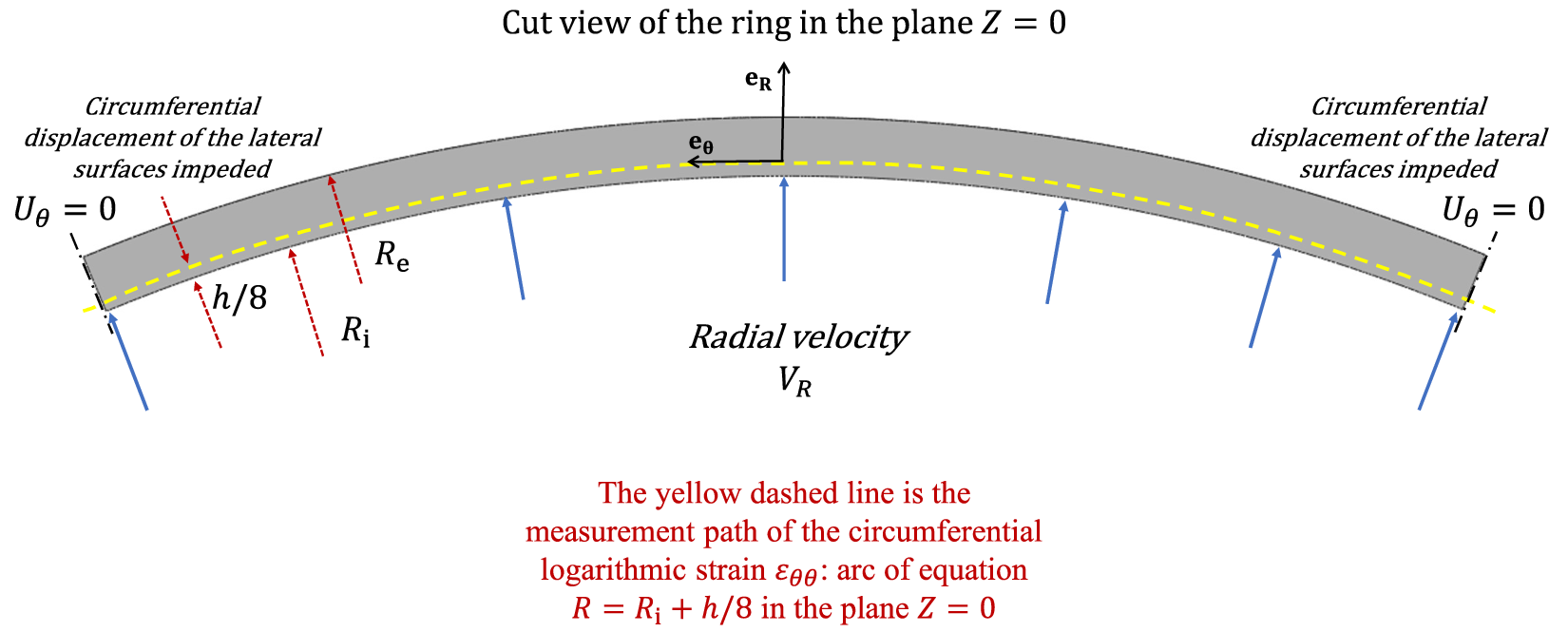}
	\label{POLYCRYSTAL_FEM_BCS}
}
\caption{Finite element model. (a) Polycrystalline microstructure N5000-D1 and mesh discretization. The origin of the cylindrical coordinate system $(R,\Theta,Z)$ is located at the center of mass of the whole specimen. The color map depicts grains with different orientations. (b) Geometry, dimensions and boundary conditions. The yellow dashed line is the measurement path of the circumferential logarithmic strain shown in Figs. \ref{ISOTROPIC_TEXTURE_RING_V_1000}, \ref{001-THETA_RING_V_1000}, \ref{001-R_TEXTURE_RING_V_1000}, \ref{111-Z_TEXTURE_RING_V_1000} and \ref{Isotropic_Slip}-\ref{Fiber_Z_Slip}.} 
\label{POLYCRYSTAL FINITE ELEMENT ELEMENT}
\end{figure}

This section describes the finite element model developed to study the formation of multiple necks in polycrystalline thin rings subjected to rapid radial expansion, see Fig. \ref{POLYCRYSTAL FINITE ELEMENT ELEMENT}. Material points are referred to using cylindrical coordinates $(R,\Theta,Z)$. The origin of the coordinate system is located at the center of mass of the ring. 

\

The inner and outer radii of the specimen are $R_{\mathrm{i}}=15~{\mathrm{mm}}$ and $R_{\mathrm{e}}=15.5~{\mathrm{mm}}$, respectively, and the axial and radial thicknesses are $e=0.5~{\mathrm{mm}}$ and $h=R_{\mathrm{e}}-R_{\mathrm{i}}$, see Fig. \ref{POLYCRYSTAL_FEM} and \ref{POLYCRYSTAL_FEM_BCS}. Similar specimen dimensions were used in the experiments performed by \cite{Zhang06}. In order to reduce the computational time, only one eighth of the ring has been analyzed, so that, to maintain the axial symmetry of the problem, the circumferential displacement of the lateral surfaces of the specimen has been impeded, see Fig. \ref{POLYCRYSTAL_FEM_BCS}. The loading condition is a radial velocity $V_R$ applied in the inner surface of the ring which remains constant during the entire analysis \citep{Rusinek07,Vadillo12,vaz2019comparative}, whereas the initial condition is a radial velocity of the same value applied to all material points \citep{MARVIMASHHADI2020103661,MARVIMASHHADI2021102999}. The application of this initial condition minimizes the propagation of waves through the radial thickness of the ring caused by the abrupt motion of the inner surface at $t = 0$, precluding instantaneous plastic localization due to the velocity loading condition \citep{Needleman91,Xue08,vaz2019comparative}. The calculations are performed for loading velocities ranging between $250~\text{m}/\text{s}$ and $5000~\text{m}/\text{s}$, leading to nominal strain rates varying from $1.66 \cdot 10^4 ~ \text{s}^{-1} \leq \dot{\varepsilon}_{0} \leq 3.33 \cdot 10^5 ~ \text{s}^{-1}$. While the largest strain rates investigated exceed the regular experimental capabilities (depending on the specimen size, the maximum strain rate attained in ring expansion laboratory experiments is $\approx 5\cdot10^4~\text{s}^{-1}$, see \cite{Zhang06} and \cite{Janiszewski12}), exploring such a wide range of loading rates helps to enlighten the interplay between inertia effects and crystallographic microstructure. Moreover, similar range of loading velocities and strain rates was considered in the numerical simulations performed by \cite{ELMAI2022104798} to investigate the effect of surface roughness on the formation of multiple necks in round bars subjected to dynamic stretching, and in the ring expansion calculations performed by \cite{Rodriguez13b} to determine the effect of strain rate on the average neck spacing in multiple necking patterns (see Section \ref{Introduction}). Note that the largest strain rates investigated are characteristic of hypervelocity impacts \citep{HASSANI2020198} and high-energy forming operations \citep{golovashchenko2013formability}.

\

The polycrystalline microstructure of the rings has been created with the open-source software NEPER \citep{quey2011}, so that the specimens have been tessellated into an aggregate of grains using random Voronoi seeds (grains are also referred to as crystals in this paper). The color map in Fig. \ref{POLYCRYSTAL_FEM} depicts the grains with different orientations. Two polycrystalline aggregates of $5000$ and $15000$ grains have been generated {(the number of grains will be denoted as $N_g$)}, and for each polycrystalline aggregate up to three different spatial distributions of seed points, D1, ..., D3, have been considered (note that each spatial distribution of seed points amounts to a different spatial distribution of grains with different orientations). The resulting microstructures are denoted by N5000-D1,..., N5000-D3, and N15000-D1, ..., N15000-D3, respectively. The idea is to investigate the effects of grain size and spatial distribution of grains on the formation of the necking pattern. Note that the microstructures with $5000$ and $15000$ grains have in average $5$ and $8$ grains {throughout} the thicknesses of the ring, which leads to an average grain size of $100~\mathrm{\mu m}$ and $62.5~\mathrm{\mu m}$, respectively. While the grains are relatively large, similar grain sizes have been reported for different commercial aluminum alloys \citep{shankar2005characteristics,Patil2014,Padmanabhan2019}. Performing calculations with smaller grain size led to excessive increase of the computational cost of the calculations (see next paragraphs). Moreover, each grain of the polycrystalline aggregate is allotted with a crystal orientation taken from one of the four different textures that have been generated with an in-house Wolfram Mathematica code developed for this purpose, namely, isotropic texture (randomly oriented grains which indeed lead to isotropic texture, see Fig. \ref{Isotropic texture pole figures}), $\left\langle 001\right\rangle\parallel\Theta$ Goss texture (see Fig. \ref{Goss texture pole figures}), $\left\langle 001\right\rangle\parallel$ R Goss texture (see Fig. \ref{110-Theta texture pole figures}) and $\left\langle 111\right\rangle\parallel$ Z fiber texture (see Fig. \ref{111-Z texture pole figures}). {The anisotropic textures were generated perturbing the corresponding ideal orientations $\bm{Q_0}$ according to the formula ($\bm{Q_0}$ and $\bm{Q_i}$ are rotation matrix transforming the local basis of the cylindrical frame $\left(R, \Theta, Z\right)$ to the crystal frame): 

\begin{eqnarray}
	\bm{Q_i}=\Delta \bm{Q_i} \left(\bm n \left(\phi_1,\phi_2 \right),\phi_3\right) \bm{Q_0} \qquad \text{with} \qquad i=1,…,N_g
	\label{Elasticity_relationship}
\end{eqnarray}

\noindent where $\Delta \bm{Q_i}$ is the rotation matrix of perturbation specified by $\bm n\left(\phi_1,\phi_2 \right)$ --the misorientation axis defined by two angles $\phi_1$, $\phi_2$ polled randomly from the intervals $\left(0,2\pi\right)$ and $\left(0,\pi\right)$, respectively, assuming uniform distribution in the interval. Moreover, $\phi_3$ is the misorientation angle also polled randomly assuming a normal (Gaussian) distribution with the mean value equal zero and $1.5^{\circ}$ standard deviation. The resulting orientation $\bm{Q_i}$ is then assigned to the grain $i$ of the ring microstructure. Recall that $N_g$ is the number of grains. All the microstructures have been generated using a specific seed (random number seed), allowing for replicable analysis.} The goal is to investigate the effect of initial texture on the formation of the necking pattern (the orientation of the crystals can change during loading). Note that the radial symmetry of the problem is broken by the different orientation of the grains, leading to perturbations in the field variables and ultimately to the formation of necks  (there is no need to introduce geometrical or material defects to break the symmetry of the problem, e.g., see \cite{Zhang10a}).

\

The ring is discretized using four node linear tetrahedral elements (C3D4 in ABAQUS notation), as they are particularly suitable to describe the complex geometry of the grains, see Fig. \ref{POLYCRYSTAL_FEM}. For the microstructures with $5000$ and $15000$ grains, the average number of elements and the standard deviation for the three distributions of seed points are $582098 \pm 1018$ and $1760760 \pm 7108$, respectively, with the average number of elements per grain being $116$ for both microstructures. Moreover, we have performed a mesh sensitivity analysis and checked that increasing the number of elements hardly affects {the strain fields, the number of necks and the necking time} predicted by the finite element simulations (the results of the mesh sensitivity analysis are not shown for the sake of brevity). {We are aware that using tetrahedral elements to discretize the ring introduces artificial stiffness in the model which may delay necking. Nevertheless, it is difficult to provide a quantitative measure of the effect that using this type of element has on the finite element results. Note that we have also used C3D4 elements to discretize rings and plates containing explicitly resolved pores and subjected to dynamic loading \citep{MARVIMASHHADI2021102999,NIETOFUENTES2022103418}, and the necking predictions were consistent with experiments and numerical calculations available in the literature, which reinforces the idea that the trends presented in this paper are hardly affected by the mesh.} The calculations have been performed with the dynamic implicit solver of \cite{ABAQUS19Standard}, using {$1$ core of} a workstation Intel Xeon Gold $6128$ @ $3.4$ GHz. {The time increment was set by the code.} The computational cost of each simulation ranged between $1$ and $17$ days, depending on the microstructure considered and the applied velocity. {Increasing the loading velocity has a twofold impact on the computational time, decreasing the necking time because the loading process is shorter (i.e., reducing the computational cost), but increasing the necking strain due to inertia effects (i.e., increasing the computational cost).} Note that the finite element results have been processed by first converting the ABAQUS ODB files to VTK format for ParaView using the Python script developed by \cite{Liu2017}.

\section{Results}
\label{Results}

The presentation of results is divided into three sections: the effects of grain size, spatial distribution of grains and texture development on the necking pattern are investigated in Section \ref{Salient features} for the initially isotropic material, Section \ref{The effect of initial texture} shows the influence of initial texture on the necking time, on the necking pattern and on the distribution of grain orientations after necking localization, and Section \ref{The effect of of impact velocity} illustrates the effect of {loading} velocity on the number of necks for different initial textures.

\subsection{Salient features}
\label{Salient features}

\begin{figure}[hbtp]
\centering
	\subfloat[]{	
		\includegraphics[width=9cm]{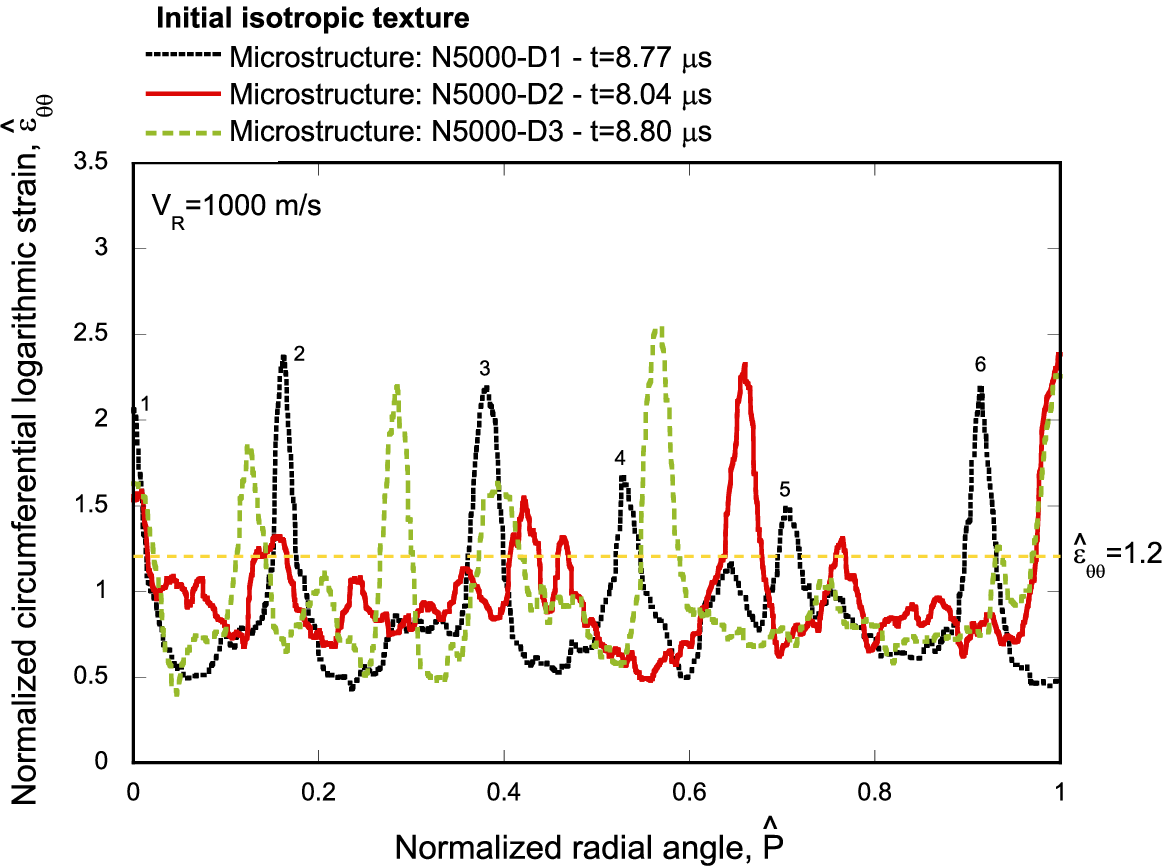}
		\label{ISOTROPIC_TEXTURE_RING_N5000_V_1000}
		}
	\subfloat[]{	
		\includegraphics[width=9cm]{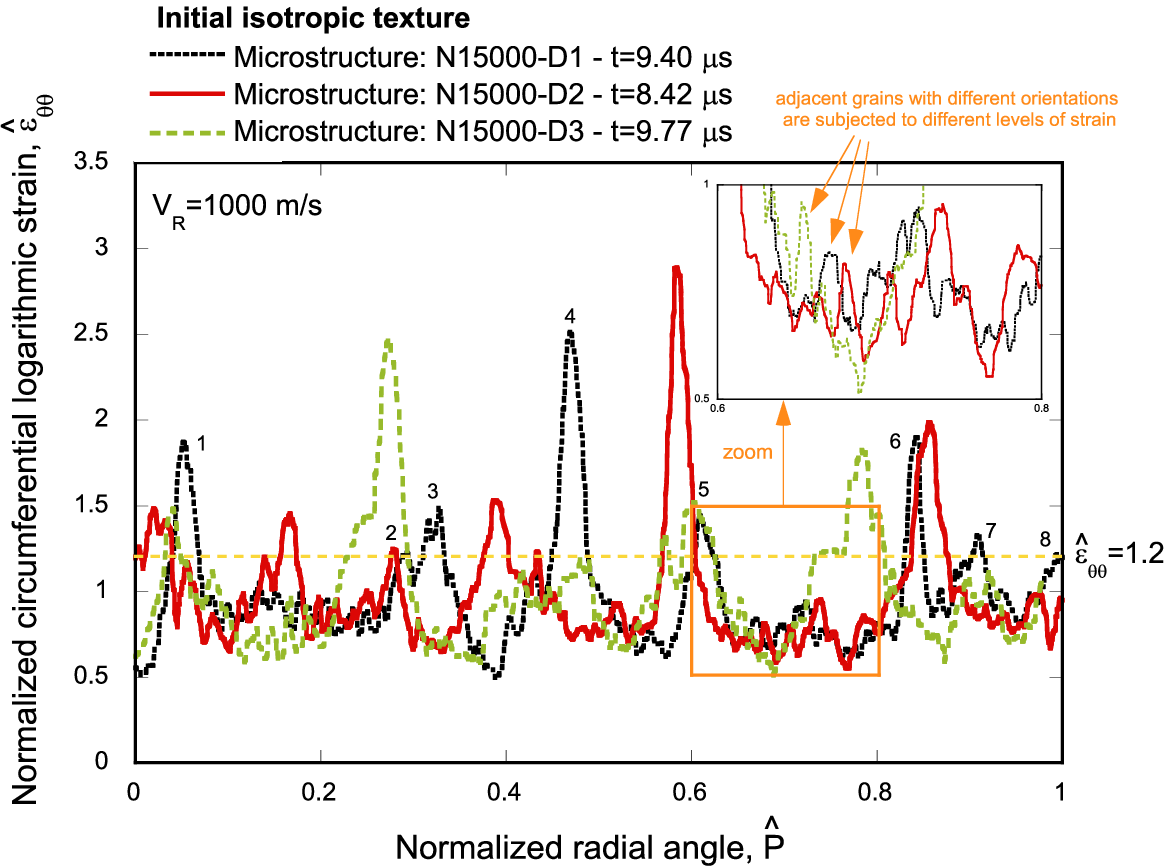}
		\label{ISOTROPIC_TEXTURE_RING_N15000_V_1000}
		}
\caption{Initial isotropic texture. Normalized circumferential logarithmic strain $\hat{\varepsilon}_{\theta\theta}$ versus normalized perimeter of the ring $\hat{P}$ for calculations performed with {loading} velocity $V_{R}=1000~\mathrm{m/s}$. Results corresponding to the necking time for microstructures: (a) N5000-D1, N5000-D2 and N5000-D3 and (b) N15000-D1, N15000-D2 and N15000-D3. The horizontal yellow dashed line corresponds to $\hat{\varepsilon}_{\theta\theta}=1.2$. For interpretation of the references to color in this figure legend, the reader is referred to the web version of this article.} 
\label{ISOTROPIC_TEXTURE_RING_V_1000}
\end{figure}

The calculations presented in this section are performed with the initially isotropic texture (grains randomly oriented), for a {loading} velocity of $V_R=1000~\text{m}/\text{s}$, which corresponds to a nominal strain rate of $\dot{\varepsilon}_0=6.66 \cdot 10^4~\text{s}^{-1}$. Fig. \ref{ISOTROPIC_TEXTURE_RING_V_1000} shows the normalized circumferential logarithmic strain $\hat{\varepsilon}_{\theta\theta}$ versus the normalized perimeter of the ring $\hat{P}=\frac{4 \Theta}{\pi}$. Recall from Section \ref{Finite element model} that only one eighth of the ring is modeled in the finite element simulations. The normalized perimeter of the ring is also referred to as normalized radial angle. The results correspond to aggregates containing $5000$ and $15000$ grains, generated with three different spatial distributions of grains: N5000-D1,..., N5000-D3, and N15000-D1, ..., N15000-D3. The notation used to designate the microstructures was introduced in Section \ref{Finite element model}. The normalized circumferential logarithmic strain is defined as $\hat{\varepsilon}_{\theta\theta}=\frac{\varepsilon_{\theta\theta}}{\varepsilon_{\theta\theta}^b}$, where $\varepsilon_{\theta\theta}$ is the circumferential logarithmic strain measured along the path shown in Fig. \ref{POLYCRYSTAL_FEM_BCS}, and $\varepsilon_{\theta\theta}^b=\ln\left(\frac{R_i+V_R t}{R_i} \right)$ approximates the background circumferential logarithmic strain in the ring, where $t$ refers to the loading time. The background strain corresponds to the fundamental theoretical solution of the problem, which is computed assuming homogeneous deformation of the specimen, so that before plastic localization occurs $\hat{\varepsilon}_{\theta\theta} \approx 1$ (see Fig. 3a in \cite{MARVIMASHHADI2020103661}). The normalized circumferential logarithmic strain facilitates to perform one-to-one comparisons of necking patterns obtained from calculations carried out at different {loading} velocities and with different initial textures, for which the formation of necks occurs at (very) different logarithmic strains (see \cite{MARVIMASHHADI2020103661}). The $\hat{\varepsilon}_{\theta\theta}-\hat{P}$ curves of Fig. \ref{ISOTROPIC_TEXTURE_RING_V_1000} correspond to the necking time and display a succession of excursions of strain which stand for sections of the ring with localized plastic deformation. Following \cite{Nsouglo2021} and \cite{jacques2021influence}, the necking time is the loading time when the condition $\dot{\bar{\varepsilon}}<10^{-3}~\text{s}^{-1}$ is met for the first time for any material point of the path indicated in Fig. \ref{POLYCRYSTAL_FEM_BCS}. Note that $\dot{\bar{\varepsilon}}=\sqrt{\frac{2}{3} \, \tilde{\mathbf{D}}^{p} : \tilde{\mathbf{D}}^{p}}$ is a scalar measure of plastic strain rate. The necking criterion corresponds to the specimen unloading, indicating that plastic strain has localized, giving rise to the formation of necking instabilities. The necking pattern is assumed to be (fully) formed at the necking time. Moreover, note that, while $10^{-3}~\text{s}^{-1}$ is an arbitrary value of plastic strain rate for the necking condition, we have checked that the necking time is largely independent of this threshold value, provided that it is small enough as compared to the nominal strain rate, so that the sections of the ring outside the excursions of strain are unloaded. Similarly to \cite{MARVIMASHHADI2020103661}, the excursions of strain that meet the condition $\hat{\varepsilon}_{\theta\theta}\geq1.2$ at the necking time are taken to be the necks of the localization pattern (the results in Figs. \ref{ISOTROPIC_TEXTURE_RING_V_1000}, \ref{001-THETA_RING_V_1000}, \ref{001-R_TEXTURE_RING_V_1000}, \ref{111-Z_TEXTURE_RING_V_1000} and \ref{Isotropic_Slip}-\ref{Fiber_Z_Slip} correspond to the necking time). {This criterion is relaxed by $5\%$ since, as the necking condition is approached, the simulations yield a field output every $0.25-1~\mu\text{s}$ (it is difficult to obtain more field and history outputs due to the increase in the size of the ODB file), which determines the error in the identification of the necking time and thus in the number of necks (the actual condition for the necking time may be met between the selected field output included in Figs. \ref{ISOTROPIC_TEXTURE_RING_V_1000}, \ref{001-THETA_RING_V_1000}, \ref{001-R_TEXTURE_RING_V_1000}, \ref{111-Z_TEXTURE_RING_V_1000} and \ref{Isotropic_Slip}-\ref{Fiber_Z_Slip}, and the next one corresponding to $\approx 0.25-1~\mu\text{s}$ later). This explains that necks $3$ and $4$ in Figs. \ref{001-THETA_RING_N5000_V_1000} and Fig \ref{001-R_TEXTURE_RING_N5000_V_1000} are taken to be necks, while the peak strain is slightly below $1.2$. (the necking time and the number of necks for all calculations reported in the paper are included in \ref{Necking time and number of necks})}. The necks for the microstructures N5000-D1 and N15000-D1 are indicated with black numbers in Figs. \ref{ISOTROPIC_TEXTURE_RING_N5000_V_1000} and \ref{ISOTROPIC_TEXTURE_RING_N15000_V_1000}, respectively. Moreover, \cite{MARVIMASHHADI2020103661} showed that this criterion to determine the number of necks is not generally very sensitive to the cut-off value of $\hat{\varepsilon}_{\theta\theta}$ chosen ($1.2$ in this paper), providing numerical predictions for the number of necks which are consistent with the experimental evidence (the reader is referred to the works of \cite{Guduru02} and \cite{ELMAI2022104798} to find alternative criteria for the identification of necks in multiple necking patterns). Note that the \textit{roughness} of the {$\hat{\varepsilon}_{\theta\theta}-\hat{P}$} curves, which show many \textit{little peaks} of strain of smaller size than the necks, see the zoomed area in Fig. \ref{ISOTROPIC_TEXTURE_RING_N15000_V_1000}, is due to the fact that adjacent grains with different orientations are subjected to different levels of strain, which leads to sharp transitions of strain at the grain boundaries. Recall from Section \ref{Introduction} that \cite{dequiedt2021localization} performed crystal plasticity calculations to study multiple necking formation in thin plates subjected to dynamic biaxial stretching. Similarly to the results shown in Fig. \ref{ISOTROPIC_TEXTURE_RING_V_1000}, \cite{dequiedt2021localization} obtained strain profiles with fluctuations of strain from grain to grain that they also attributed to differences in the local slip resistance of the grains because of their different lattice orientation.

Fig. \ref{ISOTROPIC_TEXTURE_RING_N5000_V_1000} includes $\hat{\varepsilon}_{\theta\theta}-\hat{P}$ curves for the three microstructures generated with $5000$ grains (recall that all the $\hat{\varepsilon}_{\theta\theta}-\hat{P}$ curves shown in this paper correspond to the necking time). Attending to the necking criterion (introduced in the paragraph above), $6$ necks are formed in the ring for N5000-D1 and $7$ for N5000-D2 and N5000-D3. The location of the necks varies with the spatial distribution of grains, which makes apparent the effect of the microstructure on the morphology of the localization pattern. The necking time also shows a slight variation with the microstructure, being $\approx 9 \%$ smaller for N5000-D2 than for N5000-D1 and N5000-D3 {(note that this difference lies within the error in the determination of the necking time based on the criterion given in previous paragraphs)}.

Fig. \ref{ISOTROPIC_TEXTURE_RING_N15000_V_1000} shows results for the microstructures with $15000$ grains. The $\hat{\varepsilon}_{\theta\theta}-\hat{P}$ curves are qualitatively the same than for calculations with $5000$ grains. For N15000-D1 the necking pattern consists of $8$ necks, while $7$ necks are formed for N15000-D2 and N15000-D3. While these results are not conclusive, it seems that there is {a} small increase in the number of necks with the increase in the number of grains (for the grain sizes investigated in this work). {Note that the results for the number of necks show limited statistical significance since only one-eight of the ring is modeled.} On the other hand, the necking time is $\approx 8\%$ greater in average than for the calculations with $5000$ grains, suggesting that decreasing the grain size delays the formation of the necking pattern {(yet, $8\%$ difference lies within the error in the determination of the necking time based on the criterion given in previous paragraphs)}. Note that, while the mesh is finer for the microstructures with 15000 grains, the differences in the number of necks and the necking time between the microstructures with $5000$ and $15000$ grains do not seem to be attributed to the discretization, as decreasing the element size generally leads to the opposite effect (i.e., it leads to earlier formation of necks, see \cite{NIETOFUENTES2022103418}). The results {obtained} in this paper are consistent with the calculations performed by \cite{dequiedt2021localization} with $4000$ and $50000$ grains which showed that decreasing the grain size delays necking formation in thin plates subjected to biaxial stretching. { The increasing necking time with decreasing grain size can be attributed to a decrease of the perturbation amplitude introduced by the tessellation with the increasing number of grains. This observation seems to be consistent with the results of \cite{el2014etude} --see equation (4.10) therein-- which showed that increasing the amplitude of perturbations leads to a decrease of the necking time in thin rods with imposed geometrical perturbations and subjected to dynamic stretching.}

\

\begin{figure}[hbtp]
	\centering
	\subfloat[]{	
		\includegraphics[width=15cm]{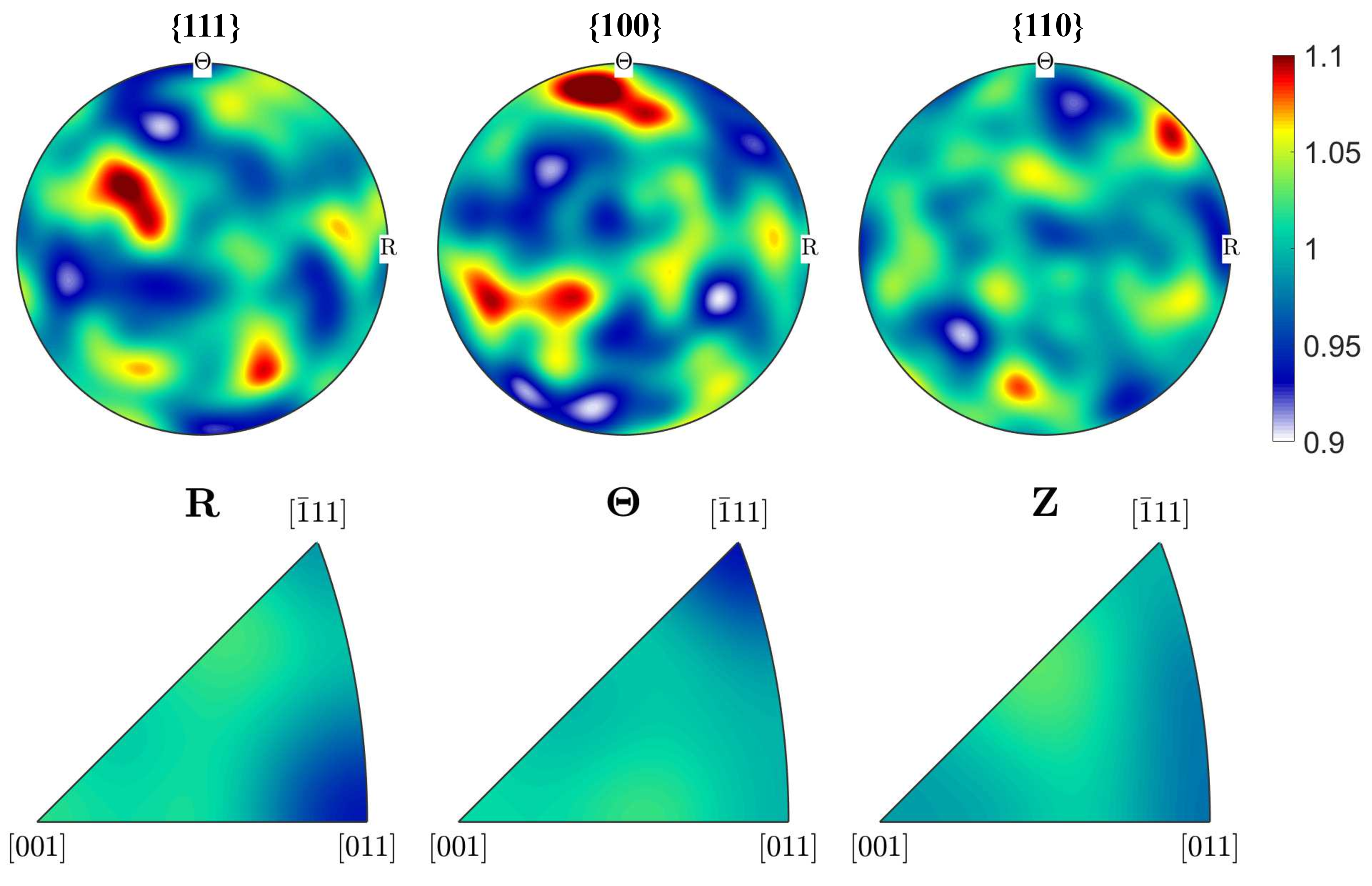}
		\label{ISOTROPIC_TEXTURE_N5000-D1_V_1000_PFs_INITIAL}
	}\\
	\subfloat[]{	
		\includegraphics[width=15cm]{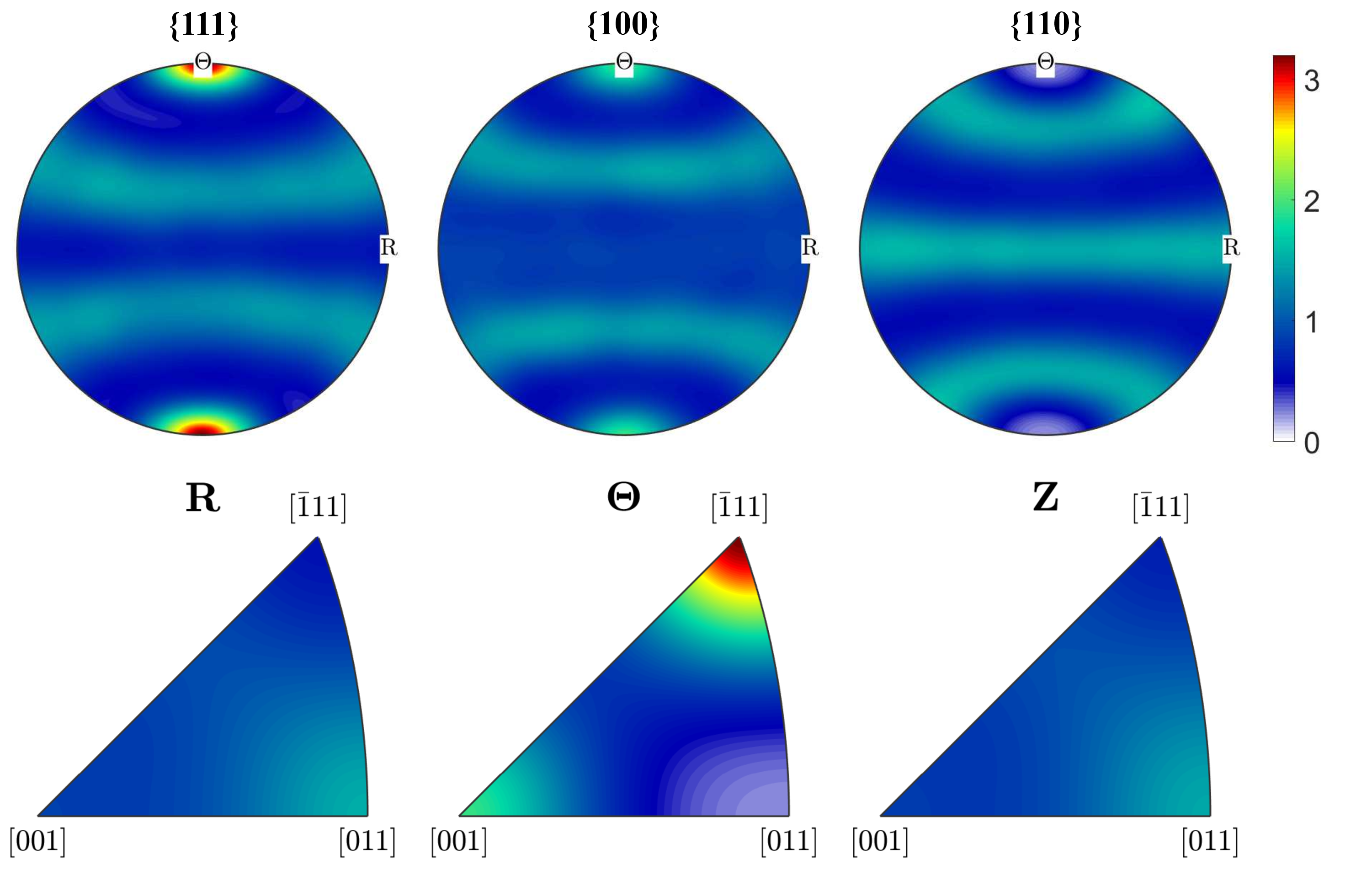}
		\label{ISOTROPIC_TEXTURE_N5000-D1_V_1000_PFs_FINAL}
	}
	\caption{Initial isotropic texture. Microstructure N5000-D1. Pole figures for $\left\lbrace 111\right\rbrace$, $\left\lbrace 100\right\rbrace $ and $\left\lbrace 110\right\rbrace$ plane families with respect to $(R,\Theta,Z)$ coordinate system, and inverse pole figures for R, $\Theta$ and Z directions with respect to $\left[001\right]$, $\left[011\right]$ and $\left[\bar{1}11\right]$ crystallographic axes. Results corresponding to: (a) undeformed and (b) deformed configurations, respectively. The deformed configuration is computed at the necking time. The calculations are performed for a {loading} velocity $V_{R}=1000~\mathrm{m/s}$. For interpretation of the references to color in this figure legend, the reader is referred to the web version of this article.} 
	\label{Isotropic texture pole figures}
\end{figure}

Texture analysis is performed for the microstructure N5000-D1. Fig. \ref{Isotropic texture pole figures} includes pole figures for $\left\lbrace 111\right\rbrace$, $\left\lbrace 100\right\rbrace $ and $\left\lbrace 110\right\rbrace$ plane families, as a standard for crystals of cubic symmetry, with respect to $(R,\Theta,Z)$ coordinate system. Inverse pole figures show orientation of local R, $\Theta$ and Z directions with respect to $\left[001\right]$, $\left[011\right]$ and $\left[\bar{1}11\right]$ crystal axes (note that by the use of the symmetry of the cubic crystal the plot is reduced to the basic triangle). All the grains of the finite element model are included in the analysis. Fig. \ref{ISOTROPIC_TEXTURE_N5000-D1_V_1000_PFs_INITIAL} shows the results for the undeformed configuration. The color coding of the isocontours is such that the texture index (measured in units of multiple of random distribution) ranges from $0.9$ to $1.1$ for a color scale that goes from white to dark red. The texture index varies within a narrow range of values, and neither the pole figures, nor the inverse pole figures show any preferential orientation of grains (as expected based on the random orientation assigned to the grains in the finite element model, so that this texture is referred to as initially isotropic). Fig. \ref{ISOTROPIC_TEXTURE_N5000-D1_V_1000_PFs_FINAL} shows the results for the deformed configuration corresponding to the necking time. The color coding is the same used for the undeformed configuration results, yet the texture index shows much wider variation from $0$ to $3.2$. The pole figures show the development of a moderate axially symmetric texture with grains that have rotated during loading to align either the crystallographic directions $\left\langle 111\right\rangle$ or the $\left\langle100\right\rangle$ with the circumferential direction of the ring. Notice that for the majority of these grains it is the $\left\langle111\right\rangle$ direction that aligns with the ring expansion direction $\Theta$, with the texture index being twice that for the $\left\langle100\right\rangle$. This is confirmed by the inverse pole figures in which $\Theta$ direction is mostly aligned with crystal axes $\left\langle 111\right\rangle$, with some amount of crystals in which it is aligned with $\left\langle 100\right \rangle$. As a consequence, the inverse pole figures for R and Z directions show that the crystallographic directions more frequently aligned with the radial and axial directions of the ring are the $\left\langle 011\right\rangle$, with texture index approximately double than the $\left\langle 111\right\rangle$ and the $\left\langle 001\right\rangle$. These results are easily explained by observing that the material volume elements along the ring circumference are locally under the state of uniaxial tension in the direction $\Theta$. For the assumed initially random orientation distribution,  deformation of these elements along the ring is then on average equivalent to the one of the extruded rod with the extrusion direction coaxial with the circumferential direction $\Theta$. Note that in the analyzed example, the inner circumference of the ring has increased to $\approxeq 18.67~\text{mm}$ by the necking time, which would correspond to the extrusion process with the rod cross-section reduction $A/A_0=0.8$ (ratio between current $A$ and initial $A_0$ cross-section areas). It is well known \citep{Kocks00} that texture developed in FCC metals in such process (if initially material was untextured) is composed of $\left\langle 111\right\rangle$ and $\left\langle100\right\rangle$ fibers with the advantage of the former one.


\subsection{The effect of initial texture}
\label{The effect of initial texture}

The results included in this section correspond to calculations with three different initial textures: $\left\langle 001\right\rangle\parallel\Theta$ Goss texture, $\left\langle 001\right\rangle\parallel$ R Goss texture and $\left\langle 111\right\rangle\parallel$ Z fiber texture. The imposed nominal strain rate is $\dot{\varepsilon}_0=6.66 \cdot 10^4~\text{s}^{-1}$ (the same used in the simulations of Section \ref{Salient features}).

\

\begin{figure}[hbtp]
	\centering
	\subfloat[]{	
		\includegraphics[width=9cm]{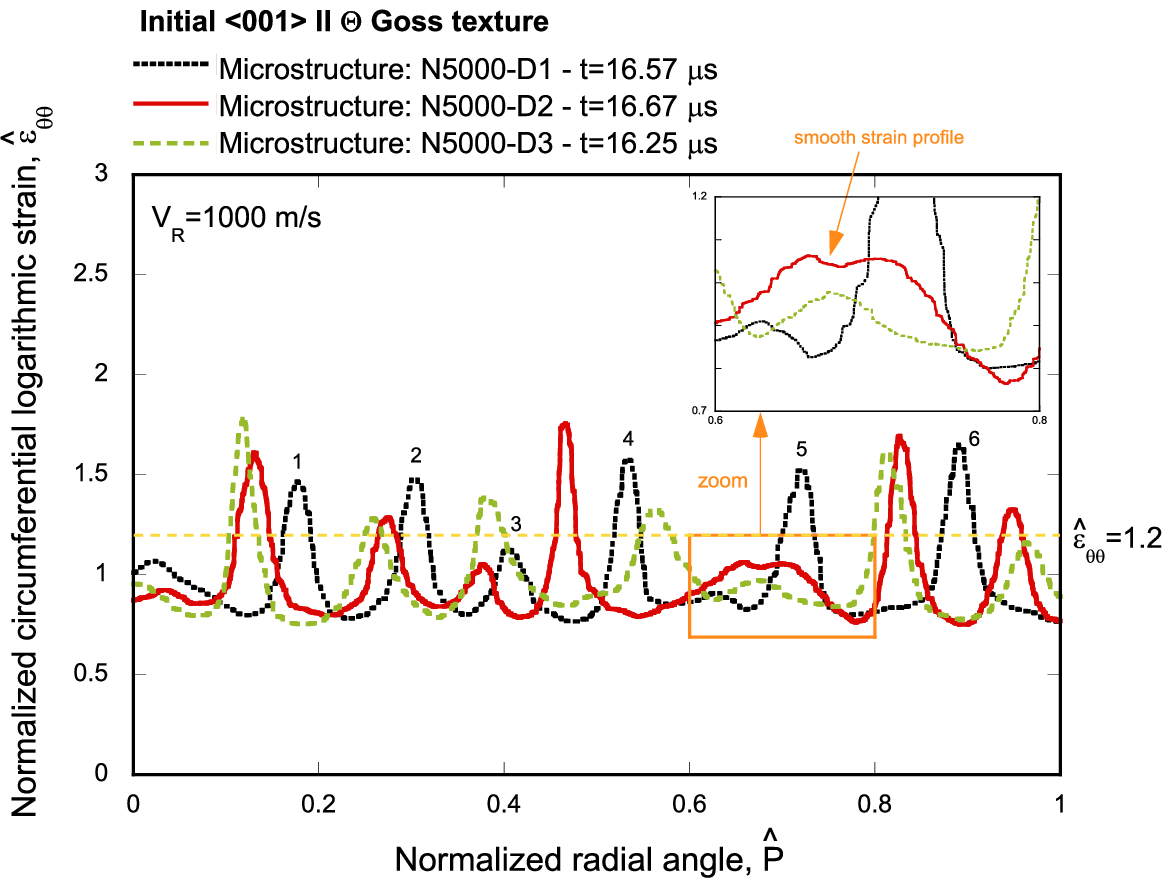}
		\label{001-THETA_RING_N5000_V_1000}
	}
	\subfloat[]{	
		\includegraphics[width=9cm]{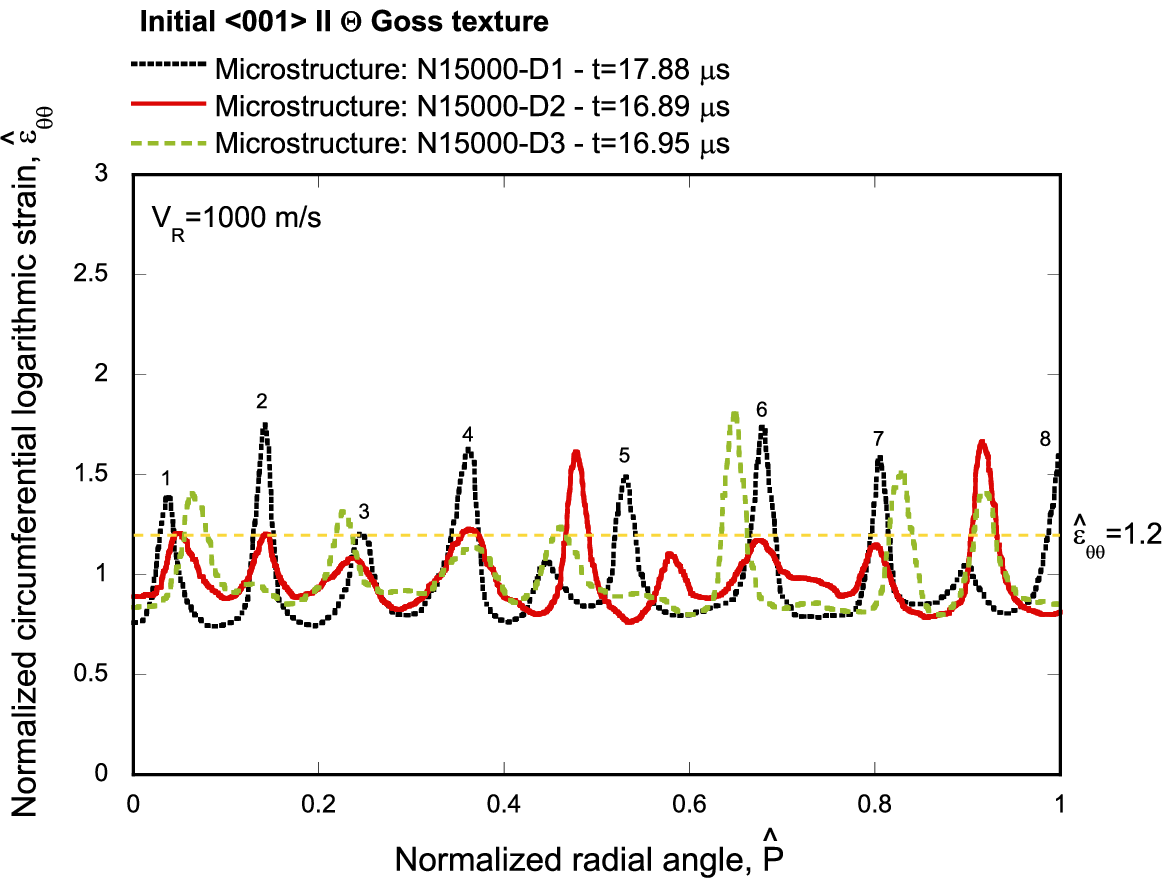}
		\label{001-THETA_RING_N15000_V_1000}
	}
	\caption{Initial $\left\langle 001\right\rangle\parallel\Theta$ Goss texture. Normalized circumferential logarithmic strain $\hat{\varepsilon}_{\theta\theta}$ versus normalized perimeter of the ring $\hat{P}$ for calculations performed with {loading} velocity $V_{R}=1000~\mathrm{m/s}$. Results corresponding to the necking time for microstructures: (a) N5000-D1, N5000-D2 and N5000-D3 and (b) N15000-D1, N15000-D2 and N15000-D3. The horizontal yellow dashed line corresponds to $\hat{\varepsilon}_{\theta\theta}=1.2$. For interpretation of the references to color in this figure legend, the reader is referred to the web version of this article.} 
	\label{001-THETA_RING_V_1000}
\end{figure}

\begin{figure}[hbtp]
	\centering
	\subfloat[]{	
		\includegraphics[width=15cm]{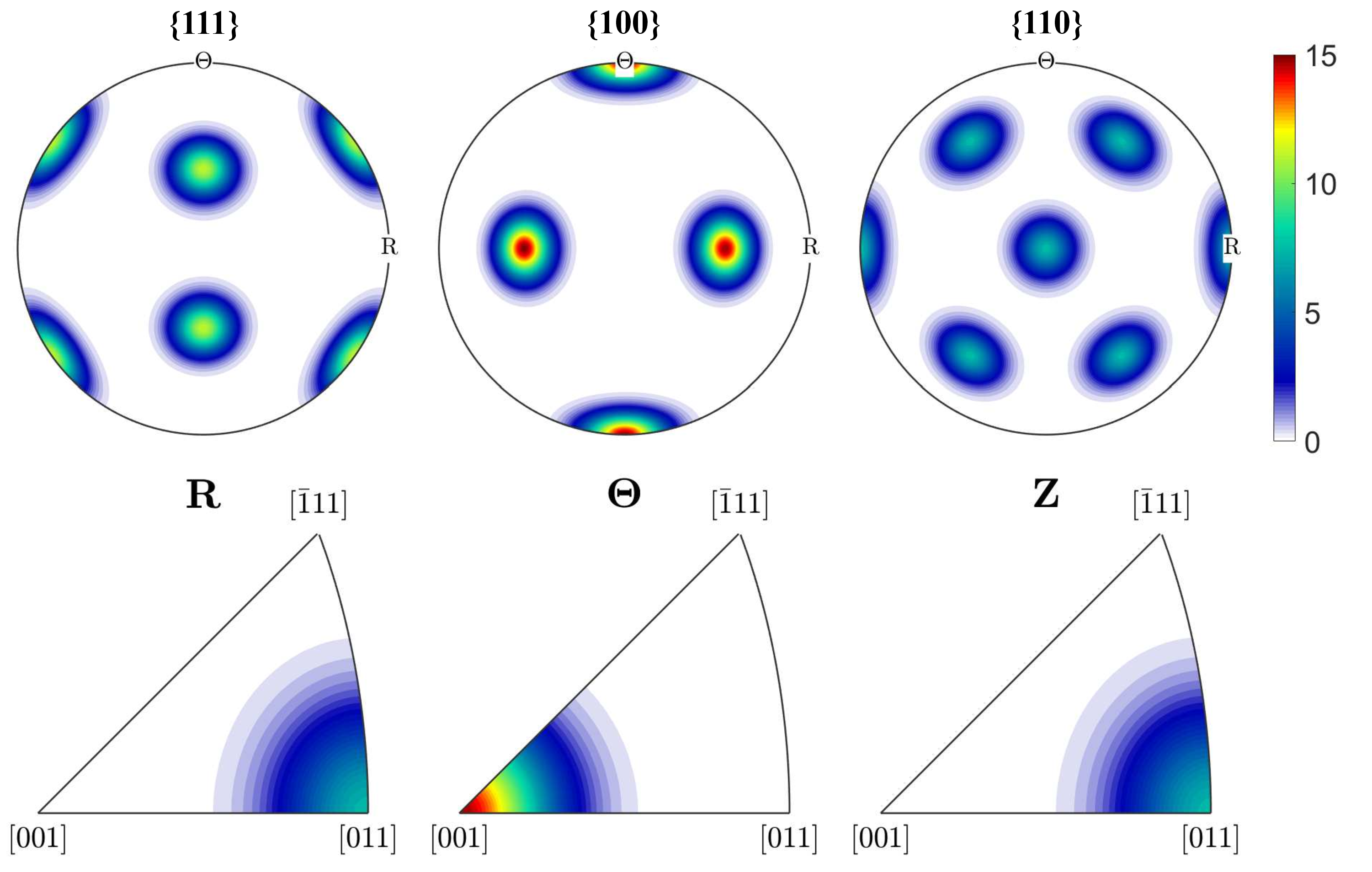}
		\label{001-THETA_N5000-D1_V_1000_PFs_INITIAL}
	}\\
	\subfloat[]{	
		\includegraphics[width=15cm]{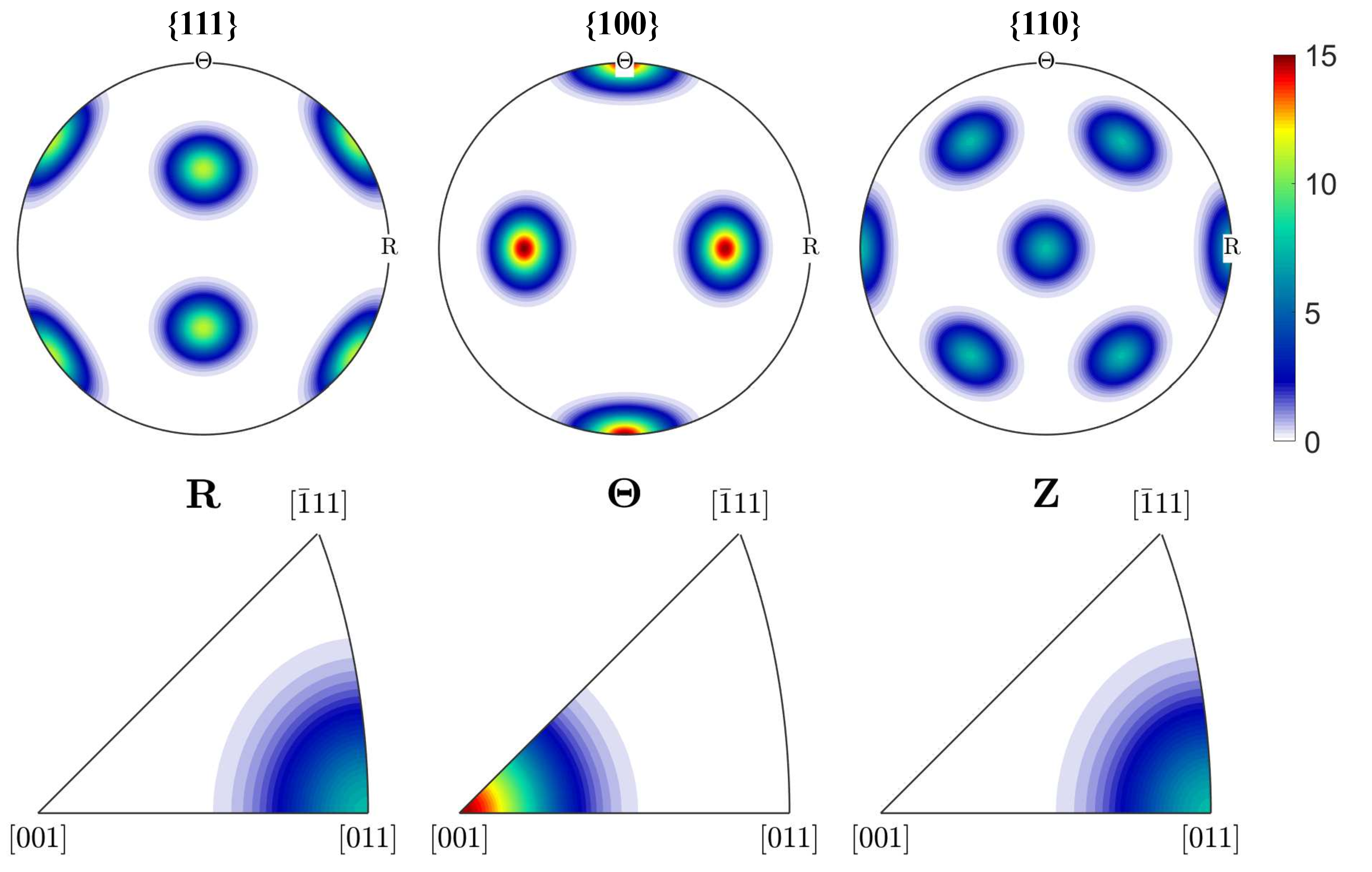}
		\label{001-THETA_N5000-D1_V_1000_PFs_FINAL}
	}
	\label{001-THETA_N5000-D1_V_1000_PFs}
	\caption{Initial $\left\langle 001\right\rangle\parallel\Theta$ Goss texture. Microstructure N5000-D1. Pole figures for $\left\lbrace 111\right\rbrace$, $\left\lbrace 100\right\rbrace $ and $\left\lbrace 110\right\rbrace$ plane families with respect to $(R,\Theta,Z)$ coordinate system, and inverse pole figures for R, $\Theta$ and Z directions with respect to $\left[001\right]$, $\left[011\right]$ and $\left[\bar{1}11\right]$ crystallographic axes. Results corresponding to: (a) undeformed and (b) deformed configurations, respectively. The deformed configuration is computed at the necking time. The calculations are performed for a {loading} velocity $V_{R}=1000~\mathrm{m/s}$. For interpretation of the references to color in this figure legend, the reader is referred to the web version of this article.} 
	\label{Goss texture pole figures}
\end{figure}

Figs. \ref{001-THETA_RING_V_1000} and \ref{Goss texture pole figures} show results for the initial $\left\langle 001\right\rangle\parallel\Theta$ Goss texture. The grains along the ring are oriented such that the crystal directions $\left\langle 001\right\rangle$ are locally parallel to the circumferential direction, while $\left\langle 110\right\rangle$ directions are aligned with the radial and axial directions (see Fig. \ref{001-THETA_N5000-D1_V_1000_PFs_INITIAL}). Selection of this texture in the analysis is driven by the fact that the Goss component $\left\langle 001\right\rangle\parallel$ RD (RD - rolling direction) is one of the components of the rolling texture in FCC metals, which if strong leads to the high in-plane anisotropy of the sheet, cf. \cite{Kocks00}. Depending on the process by which the ring is obtained from the rolled sheet one may expect components $\left\langle 001\right\rangle\parallel \Theta$ or $\left\langle 001\right\rangle\parallel$ R (analyzed in the next part of this subsection)  to be present in the ring. Note that the presence of this component after rolling process can be enhanced by some alloying additions or heat treatment.

The evolution of the normalized circumferential logarithmic strain $\hat{\varepsilon}_{\theta\theta}$ along the normalized perimeter of the ring $\hat{P}$ is shown in Figs. \ref{001-THETA_RING_N5000_V_1000} and \ref{001-THETA_RING_N15000_V_1000} for calculations performed with microstructures N5000-D1,..., N5000-D3 and N15000-D1,..., N15000-D3, respectively. The results correspond to the necking time (the criterion to determine the necking time was given in Section \ref{Salient features}). The shape of the $\hat{\varepsilon}_{\theta\theta} - \hat{P}$ curves obtained for all the microstructures is very similar regardless the number and the spatial distribution of grains. On the other hand, the specific location of the necks varies with the microstructure. The number of necks ranges between $5$ and $6$ for the calculations with $5000$ grains, and between $7$ and $8$ for the calculations with $15000$ grains. These results suggest that the number of necks for the initial $\left\langle 001\right\rangle\parallel\Theta$ Goss texture tends to increase with the decrease of the average grain size {(recall that the results for the number of necks show limited statistical significance since only one-eight of the ring is modeled)}. The necking time is also slightly larger for the aggregates with $15000$ grains, consistent with the trend obtained from the simulations with initially isotropic texture in which decreasing the grain size delayed necking formation. Moreover, notice that the necking time is almost double than for the initially isotropic texture (compare the necking times reported in Figs. \ref{ISOTROPIC_TEXTURE_RING_V_1000} and \ref{001-THETA_RING_V_1000}). This is an important finding of this investigation, which shows that texture can be tailored to delay dynamic necking formation and thus to improve the energy absorption capacity of ductile metallic materials at high strain rates. Notice also that the $\hat{\varepsilon}_{\theta\theta} - \hat{P}$ curves present much less \textit{roughness} than the calculations with initially isotropic texture (compare Figs. \ref{ISOTROPIC_TEXTURE_RING_V_1000} and \ref{001-THETA_RING_V_1000}). The strain profiles are smooth, with \textit{well-defined} strain excursions, and the reason is that the initial texture does not evolve during loading, all crystals show similar slip resistance, which leads to gentle transitions of strain from grain to grain (this conclusion is further substantiated with the texture analysis presented below).

The results of the texture analysis performed for the microstructure N5000-D1 are shown in Figs. \ref{001-THETA_N5000-D1_V_1000_PFs_INITIAL} and \ref{001-THETA_N5000-D1_V_1000_PFs_FINAL} for the undeformed configuration and for the necking time, respectively. As for the case of the initially isotropic texture, the pole figures correspond to $\left\lbrace 111\right\rbrace$, $\left\lbrace 100\right\rbrace $ and $\left\lbrace 110\right\rbrace$ plane families with respect to $(R,\Theta,Z)$ coordinate system, and the inverse pole figures to R, $\Theta$ and Z directions with respect to $\left[001\right]$, $\left[011\right]$ and $\left[\bar{1}11\right]$ crystallographic axes (we use the same representations of pole figures and inverse pole figures for all the texture analyses performed in this paper). The texture index ranges from $0$ to $15$ for a color scale that goes from white to dark red. The large gradients in the texture index shown in the pole figures illustrate the strong texture of the material, the multiple of random orientation reaching $15$ for the $\left\lbrace 100\right\rbrace$ plane families, and $11$ and $7.4$ for the $\left\lbrace 111\right\rbrace$ and $\left\lbrace 110\right\rbrace$ plane families, respectively. This is the result of the assumed orientation distribution which only slightly may deviate from the ideal configuration as described in Section \ref{Finite element model}. Notice that the texture does not change during loading (as anticipated in previous paragraph), i.e., the orientation of the crystals stays the same during the whole loading process (compare Figs. \ref{001-THETA_N5000-D1_V_1000_PFs_INITIAL} and \ref{001-THETA_N5000-D1_V_1000_PFs_FINAL}). The inverse pole figures show that the crystallographic direction $\left[011\right]$ stays aligned with the radial and axial directions of the ring, and the $\left[001\right]$ direction with the circumferential (loading) direction. These results are consistent with the analysis of the initial isotropic texture, which showed the lattice rotating during loading to align the $\left\langle100\right\rangle$ crystallographic directions with the circumferential direction of the ring in some fraction of grains (see Fig. \ref{Isotropic texture pole figures}). Note that the Goss texture $\left\langle 001\right\rangle|| \Theta$ is the component of the fiber $\left\langle 100 \right\rangle$. The latter is a stable orientation for the deformation path considered in the analysis (being in principle extrusion deformation in circumferential direction), for which, due to the lack of the lattice spin, we do not observe reorientation. For a strict definition of the stable orientation one may refer to \cite{Li08I}.

\

\begin{figure}[hbtp]
	\centering
	\subfloat[]{	
		\includegraphics[width=9cm]{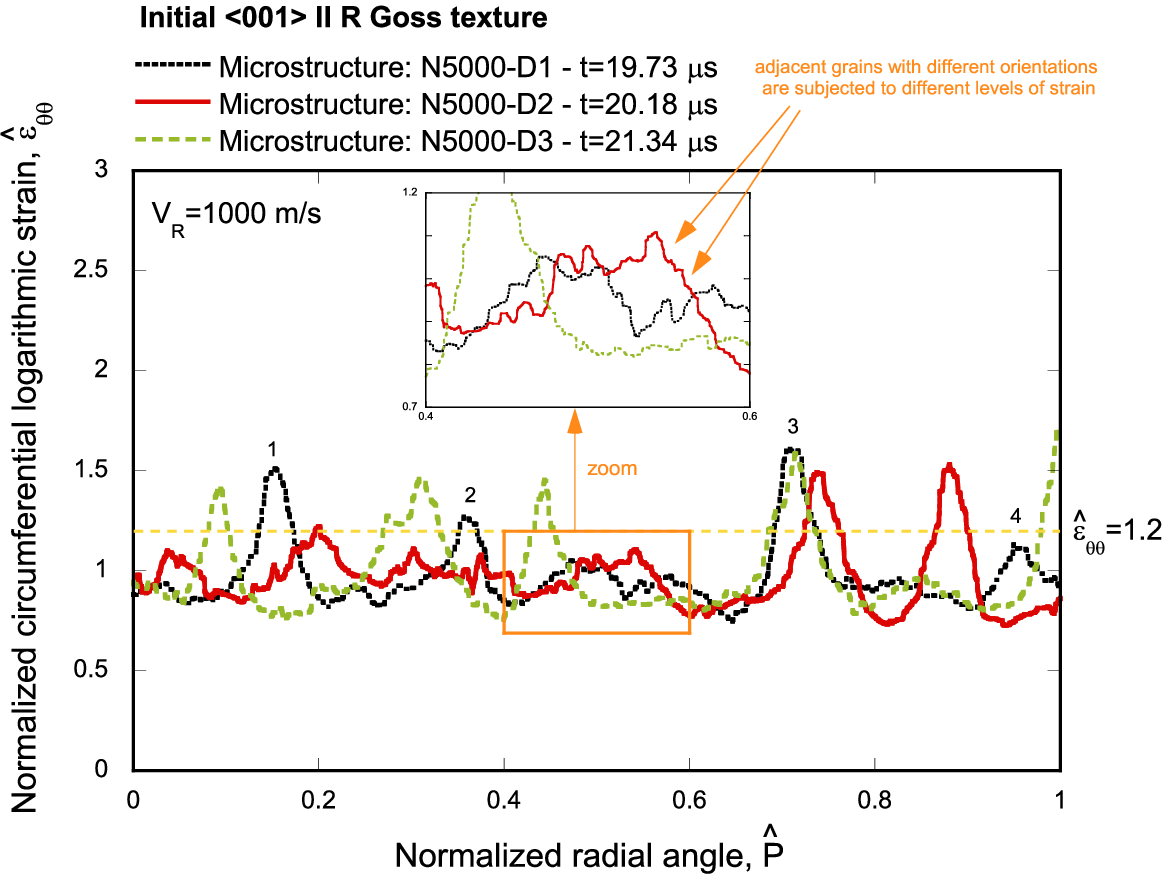}
		\label{001-R_TEXTURE_RING_N5000_V_1000}
	}
	\subfloat[]{	
		\includegraphics[width=9cm]{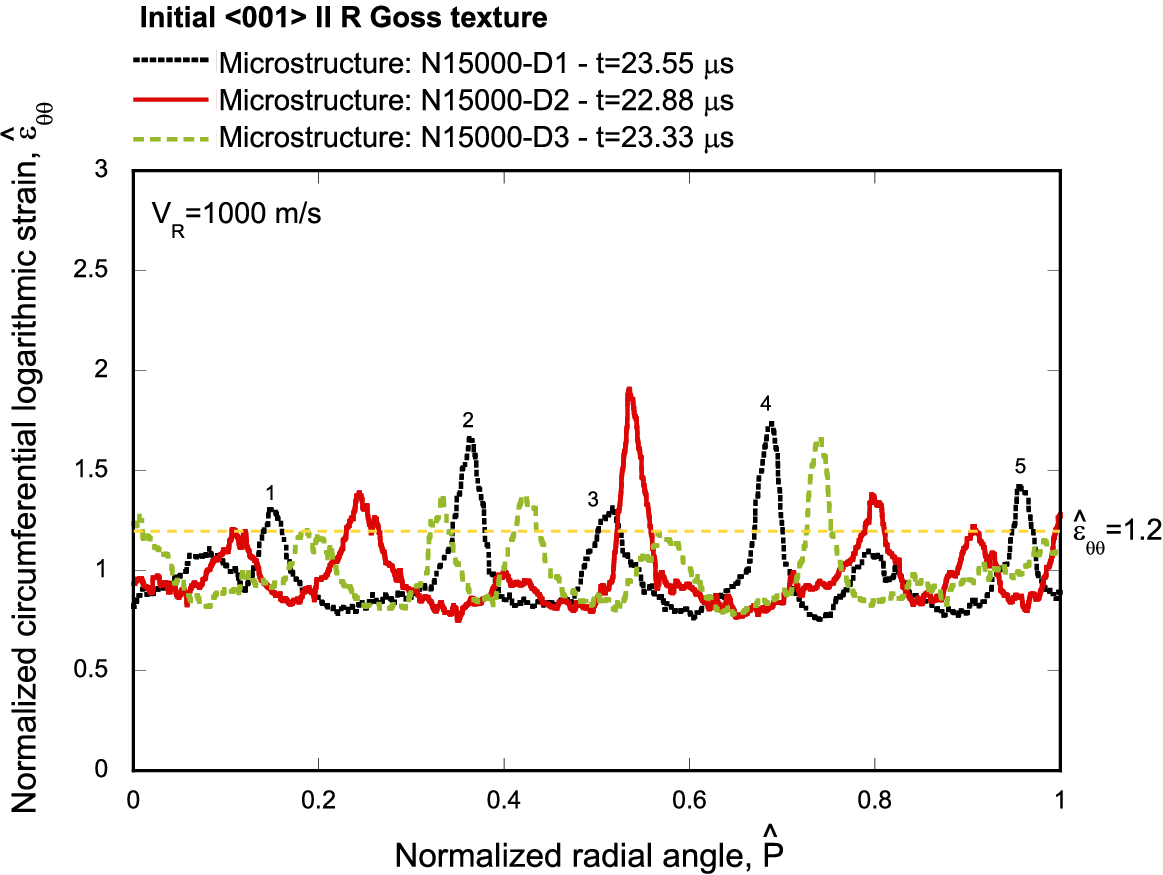}
		\label{001-R_TEXTURE_RING_N15000_V_1000}
	}
	\caption{Initial $\left\langle 001\right\rangle\parallel$ R Goss texture. Normalized circumferential logarithmic strain $\hat{\varepsilon}_{\theta\theta}$ versus normalized perimeter of the ring $\hat{P}$ for calculations performed with {loading} velocity $V_{R}=1000~\mathrm{m/s}$. Results corresponding to the necking time for microstructures: (a) N5000-D1, N5000-D2 and N5000-D3 and (b) N15000-D1, N15000-D2 and N15000-D3. The horizontal yellow dashed line corresponds to $\hat{\varepsilon}_{\theta\theta}=1.2$. For interpretation of the references to color in this figure legend, the reader is referred to the web version of this article.} 
	\label{001-R_TEXTURE_RING_V_1000}
\end{figure}

\begin{figure}[hbtp]
	\centering
	\subfloat[]{	
		\includegraphics[width=15cm]{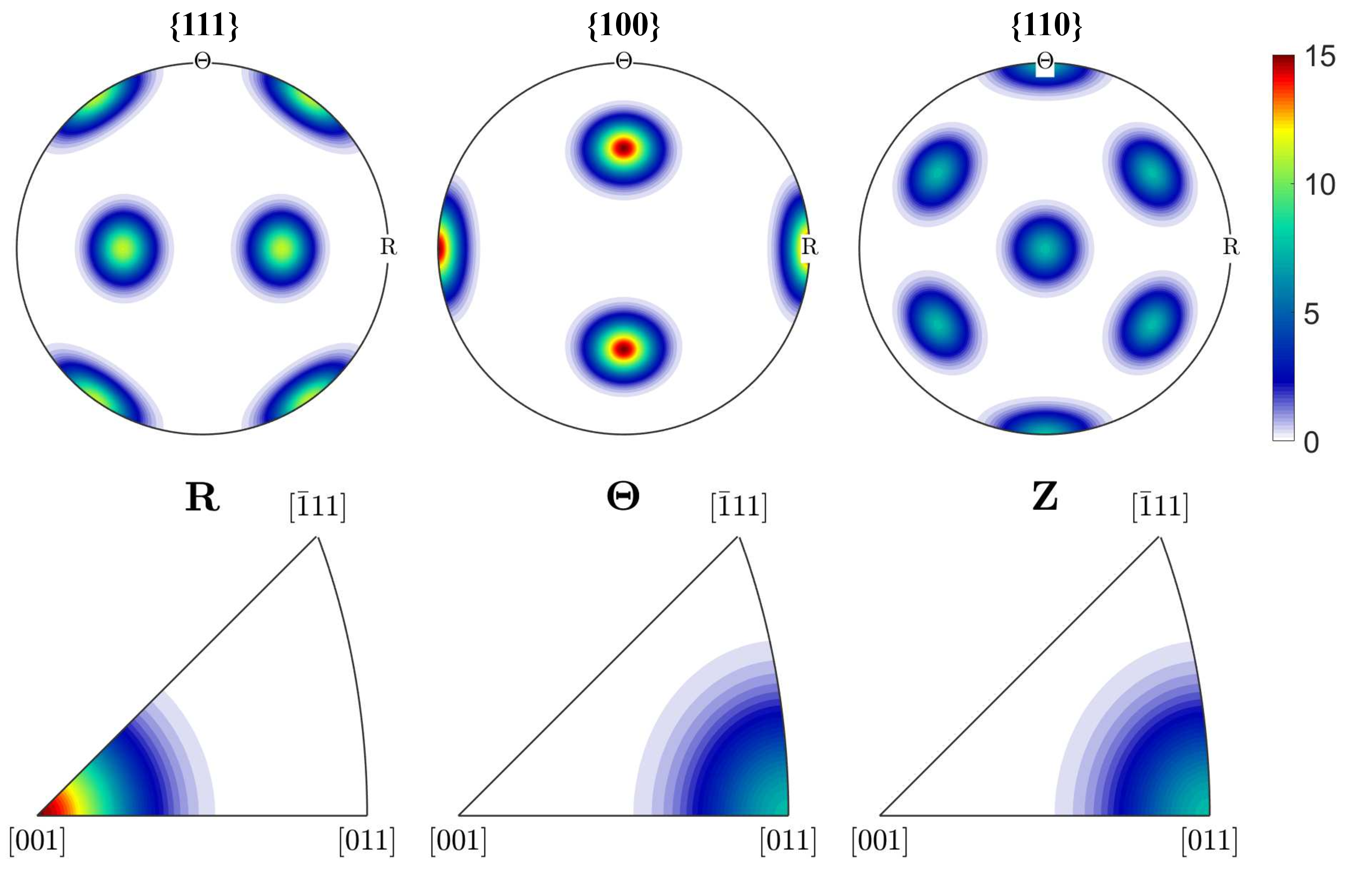}
		\label{001-R_TEXTURE_N5000-D1_V_1000_PFs_INITIAL}
	}\\
	\subfloat[]{	
		\includegraphics[width=15cm]{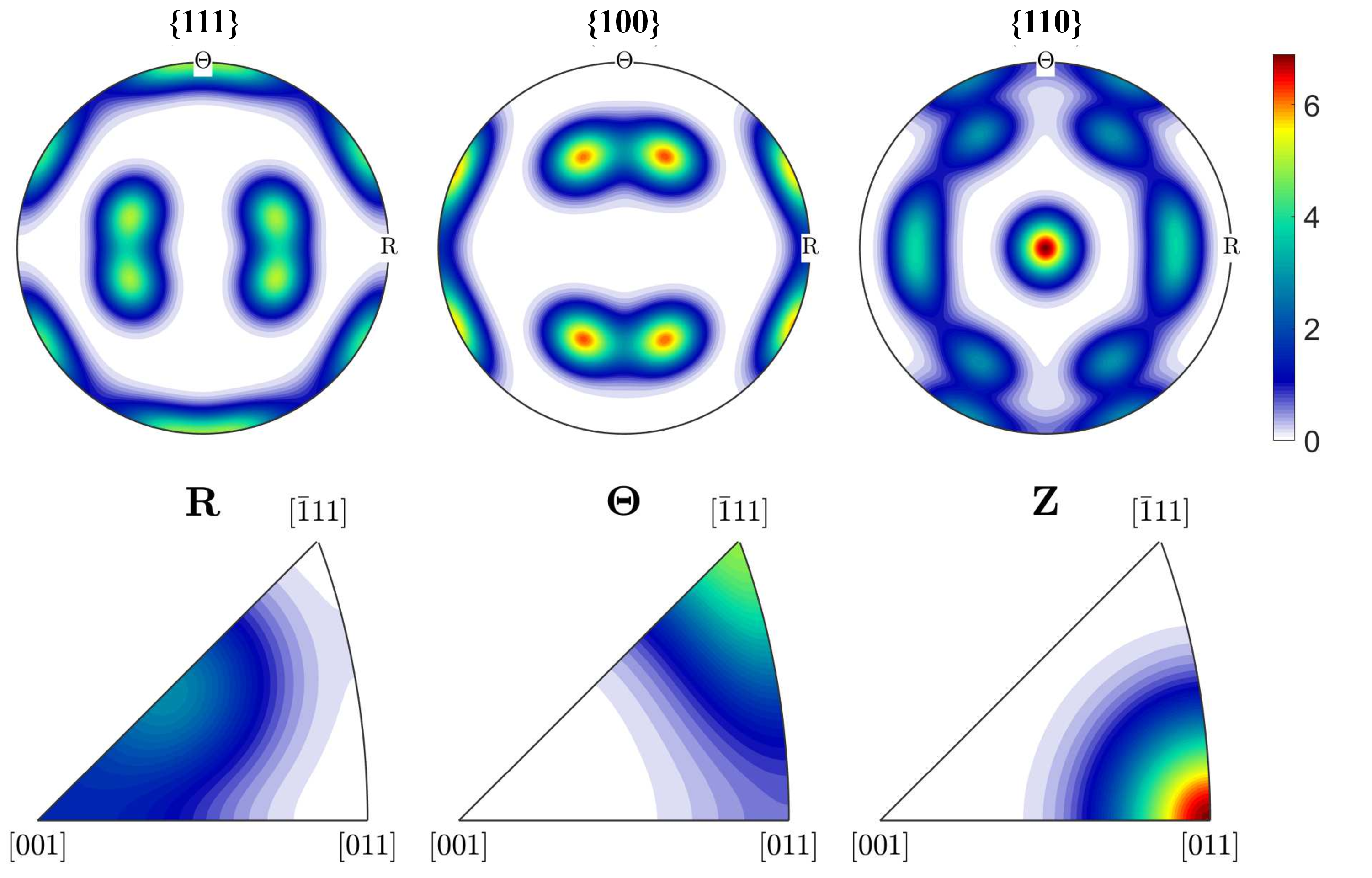}
		\label{001-R_TEXTURE_N5000-D1_V_1000_PFs_FINAL}
	}	
	\caption{Initial $\left\langle 001\right\rangle\parallel$ R Goss texture. Microstructure N5000-D1. Pole figures for $\left\lbrace 111\right\rbrace$, $\left\lbrace 100\right\rbrace $ and $\left\lbrace 110\right\rbrace$ plane families with respect to $(R,\Theta,Z)$ coordinate system, and inverse pole figures for R, $\Theta$ and Z directions with respect to $\left[001\right]$, $\left[011\right]$ and $\left[\bar{1}11\right]$ crystallographic axes. Results corresponding to: (a) undeformed and (b) deformed configurations, respectively. The deformed configuration is computed at the necking time. The calculations are performed for a {loading} velocity $V_{R}=1000~\mathrm{m/s}$. For interpretation of the references to color in this figure legend, the reader is referred to the web version of this article.} 
	\label{110-Theta texture pole figures}
\end{figure}

Figs. \ref{001-R_TEXTURE_RING_V_1000} and \ref{110-Theta texture pole figures} include results for the initial $\left\langle 001\right\rangle\parallel$ R Goss texture. The difference with respect to the previously considered $\left\langle 001\right\rangle\parallel \Theta$  texture is that the crystal directions $\left\langle 001\right\rangle$ are locally parallel to the radial ring direction, while directions $\left\langle 110\right\rangle$ are parallel to the circumferential and axial directions, so that the crystals are rotated by $90^{\circ}$ around Z axis as compared to the previous case (see Figure \ref{001-R_TEXTURE_RING_N5000_V_1000}).

The evolution of the normalized circumferential logarithmic strain $\hat{\varepsilon}_{\theta\theta}$ along the normalized perimeter of the ring $\hat{P}$ is shown in Figs. \ref{001-R_TEXTURE_RING_N5000_V_1000} and \ref{001-R_TEXTURE_RING_N15000_V_1000} for calculations performed with polycrystalline aggregates containing $5000$ and $15000$ grains, respectively. The general trends on the effect of grain size on the necking pattern are the same obtained for the initial isotropic texture and for the initial $\left\langle 001\right\rangle\parallel\Theta$ Goss texture, so that decreasing the grain size delays the necking time and leads to the formation of more necks. Namely, the difference in the average necking time is $\approx 13\%$. Moreover, the number of necks for the microstructures with $5000$ grains varies from $3$ to $5$, while for the aggregates with $15000$ grains ranges from $5$ to $7$ {(the large relative variability in the number of necks obtained in the realizations with the same number of grains is caused by the limited statistical significance of the number of necks since only one-eight of the ring is modeled)}. On the other hand, note that the $\hat{\varepsilon}_{\theta\theta}-\hat{P}$ curves show a \textit{rough} profile, similar to that of the simulations with initial isotropic texture (compare Figs. \ref{ISOTROPIC_TEXTURE_RING_V_1000} and \ref{001-R_TEXTURE_RING_V_1000}), which is caused by the evolution of the texture during loading (same reason than in the case of the initial isotropic texture). The reorientation of the crystals leads to an irregular localization pattern, and several sections of the ring with localized plastic deformation do not develop into necks (based on the necking criterion introduced in Section \ref{Salient features}). Moreover, note that the necking pattern is less uniform than in the case of the $\left\langle 001\right\rangle\parallel\Theta$ Goss texture. For instance, the average normalized peak strain of the necks and the corresponding standard deviation for the microstructure N15000-D1 in the case of initial $\left\langle 001\right\rangle\parallel\Theta$ Goss texture is $1.55\pm0.17$, while in the case of $\left\langle 001\right\rangle\parallel$ R Goss texture the peak strain decreases and the standard deviation increases $1.48\pm0.20$ {(the average normalized peak strain is the average of the normalized maximum strain of all the necks identified in the corresponding $\hat{\varepsilon}_{\theta\theta}-\hat{P}$ curve)}. Note also that the necking time is significantly greater for the $\left\langle 001\right\rangle\parallel$ R Goss texture than for the $\left\langle 001\right\rangle\parallel\Theta$ Goss texture and for the initially isotropic texture, namely, $35\%$ and $150\%$, respectively (these results are obtained considering the average necking time of the three aggregates with $15000$ grains).

The texture analysis performed for the microstructure N5000-D1 is shown in Fig. \ref{110-Theta texture pole figures}. The texture index ranges from $0$ to $15$ in the undeformed configuration, see Fig. \ref{001-R_TEXTURE_N5000-D1_V_1000_PFs_INITIAL}, and from $0$ to $6.9$ at the necking time, see Fig. \ref{001-R_TEXTURE_N5000-D1_V_1000_PFs_FINAL}, showing a decrease in the texture intensity upon deformation due to the rotation of the crystals during the expansion of the ring. At the necking time, the colored areas of the pole figures are larger, showing grains with orientation other than the initial. For instance, the inverse pole figures bring to light the tendency of the grains to align the $\left\langle111\right\rangle$ crystallographic directions with the circumferential direction of the ring, similarly to the results obtained from the texture analysis on the initially isotropic rings. Notice that, due to the introduced disturbance of the ideal orientation, the initial orientations of the crystals are not stable for the considered loading process, so that during the ring expansion the lattice rotates towards the stability point, so to orient the $\left\langle 111\right\rangle$ with circumferential direction, while keeping the $\left\langle 011\right\rangle$ directions oriented with the axial direction. This leads to the appearance of the brass-type components in the texture image for the deformed configuration. It should be noted that for the present texture, due to the cross-section anisotropy, uniaxial tension in the circumferential direction is not equivalent to the extrusion process, contrary to previous two cases.

\

\begin{figure}[hbtp]
	\centering
	\subfloat[]{	
		\includegraphics[width=9cm]{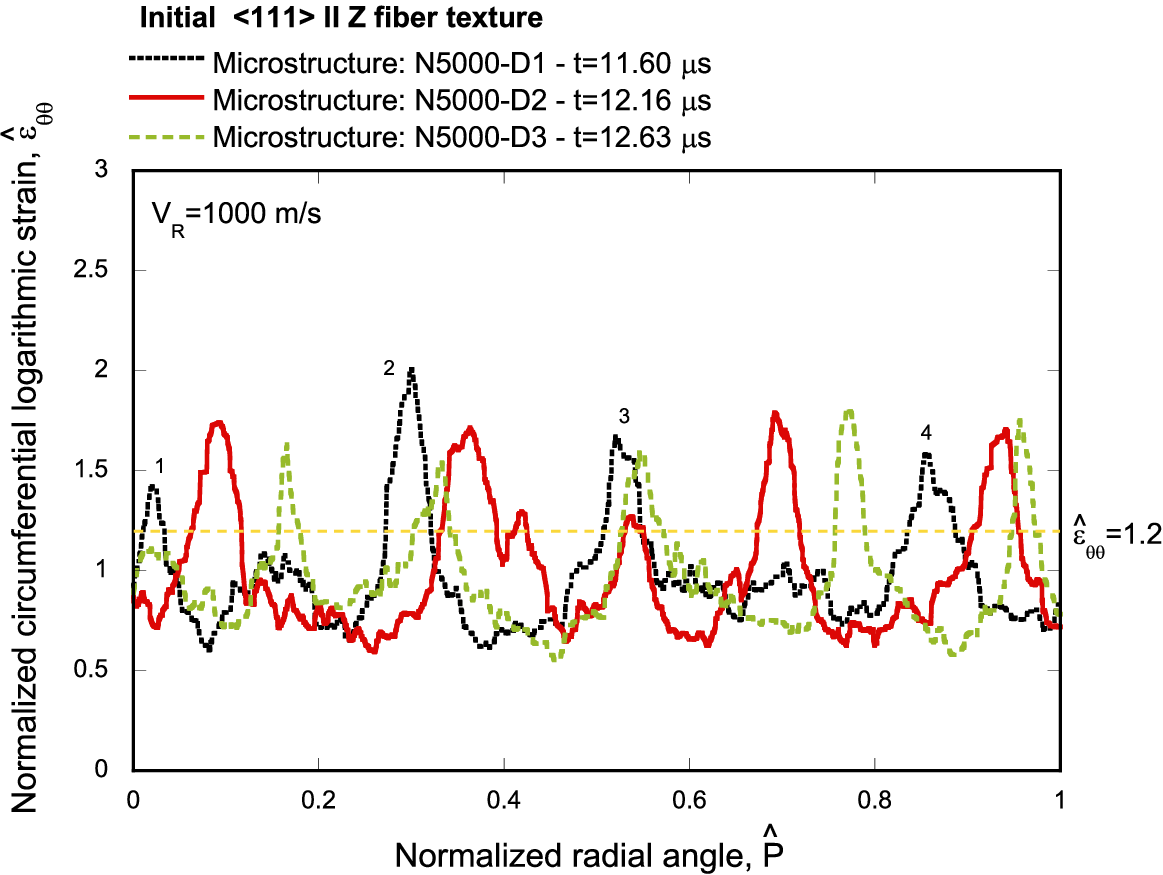}
		\label{111-Z_TEXTURE_RING_N5000_V_1000}
	}
	\subfloat[]{	
		\includegraphics[width=9cm]{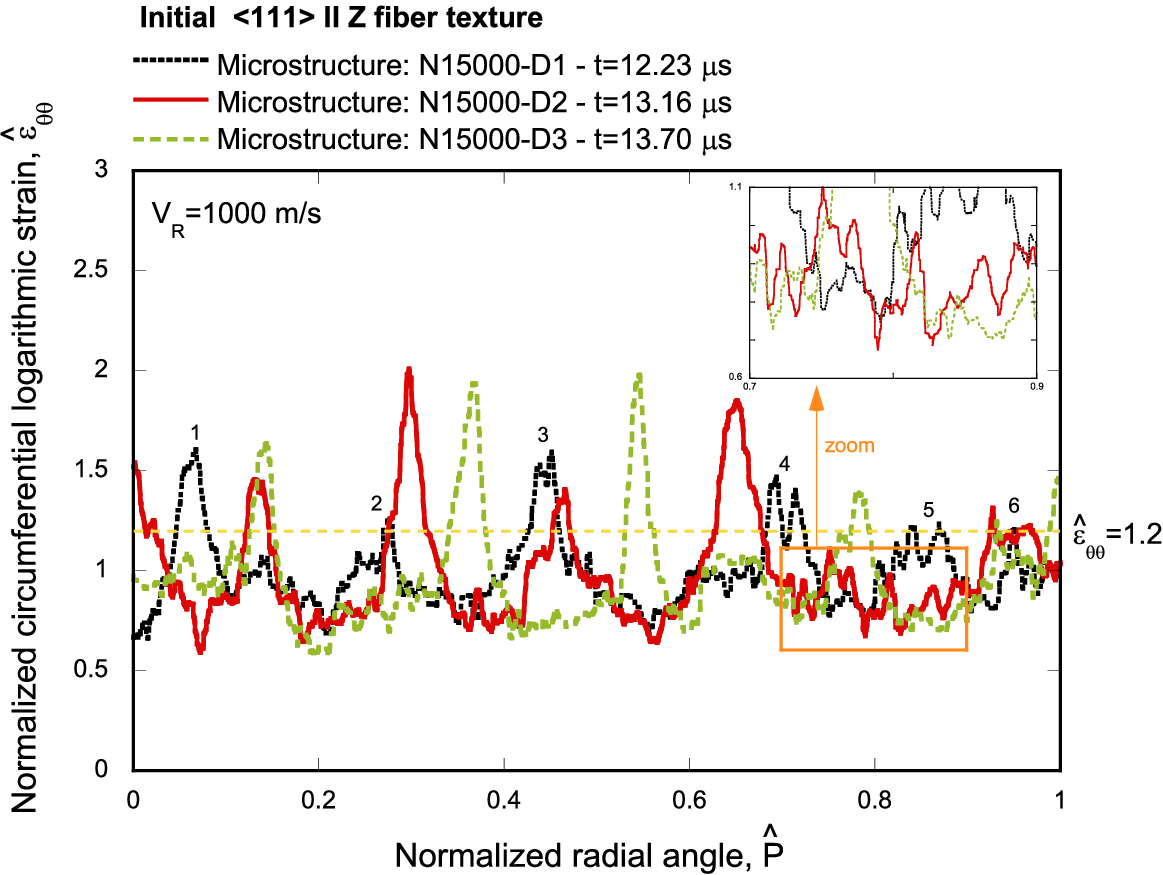}
		\label{111-Z_TEXTURE_RING_N15000_V_1000}
	}
	\caption{Initial $\left\langle 111\right\rangle\parallel$ Z fiber texture. Normalized circumferential logarithmic strain $\hat{\varepsilon}_{\theta\theta}$ versus normalized perimeter of the ring $\hat{P}$ for calculations performed with {loading} velocity $V_{R}=1000~\mathrm{m/s}$. Results corresponding to the necking time for microstructures: (a) N5000-D1, N5000-D2 and N5000-D3 and (b) N15000-D1, N15000-D2 and N15000-D3. The horizontal yellow dashed line corresponds to $\hat{\varepsilon}_{\theta\theta}=1.2$. For interpretation of the references to color in this figure legend, the reader is referred to the web version of this article.} 
	\label{111-Z_TEXTURE_RING_V_1000}
\end{figure}

\begin{figure}[hbtp]
	\centering
	\subfloat[]{	
		\includegraphics[width=15cm]{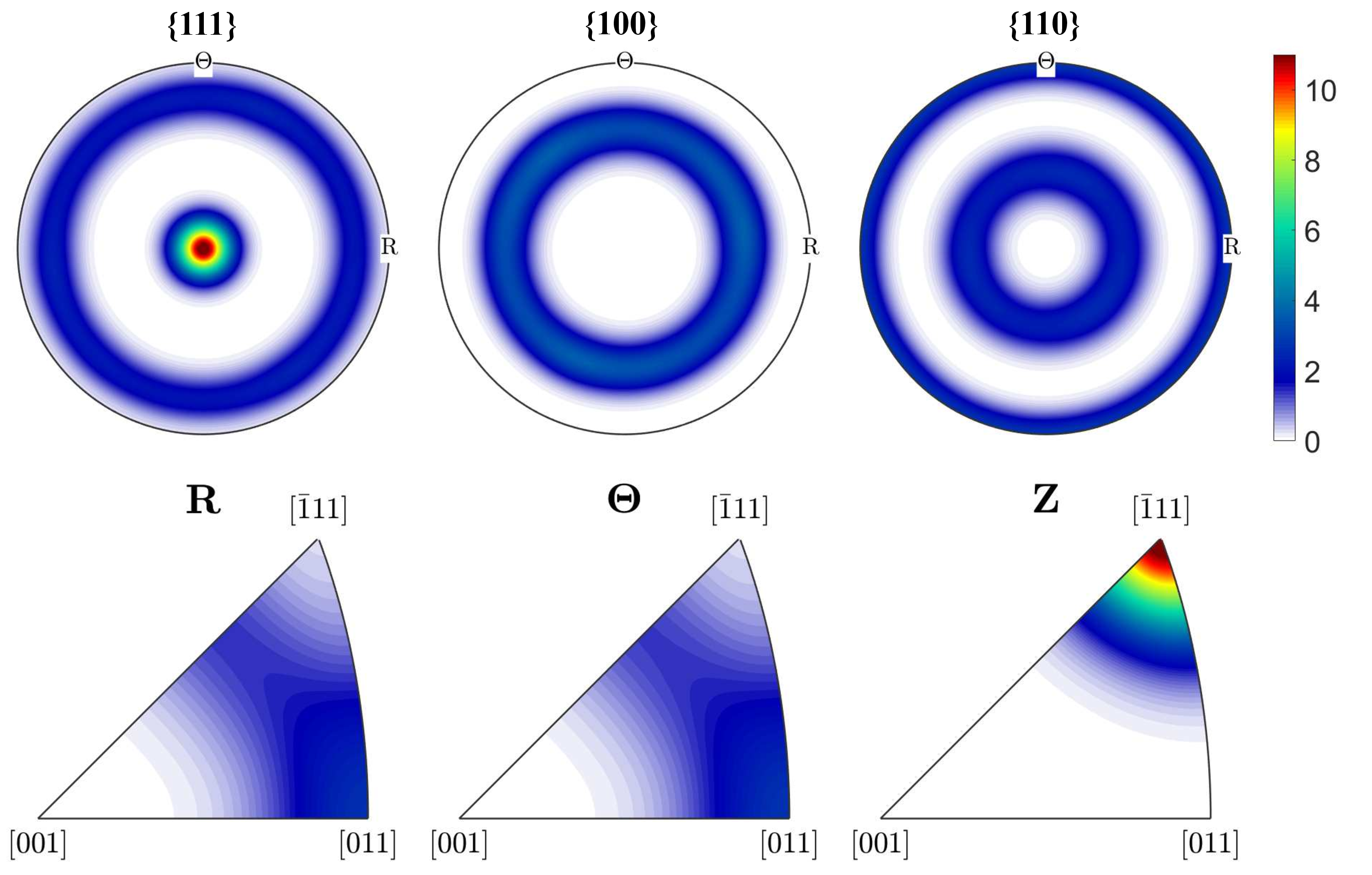}
		\label{111-Z_TEXTURE_N5000-D1_V_1000_PFs_INITIAL}
	}\\
	\subfloat[]{	
		\includegraphics[width=15cm]{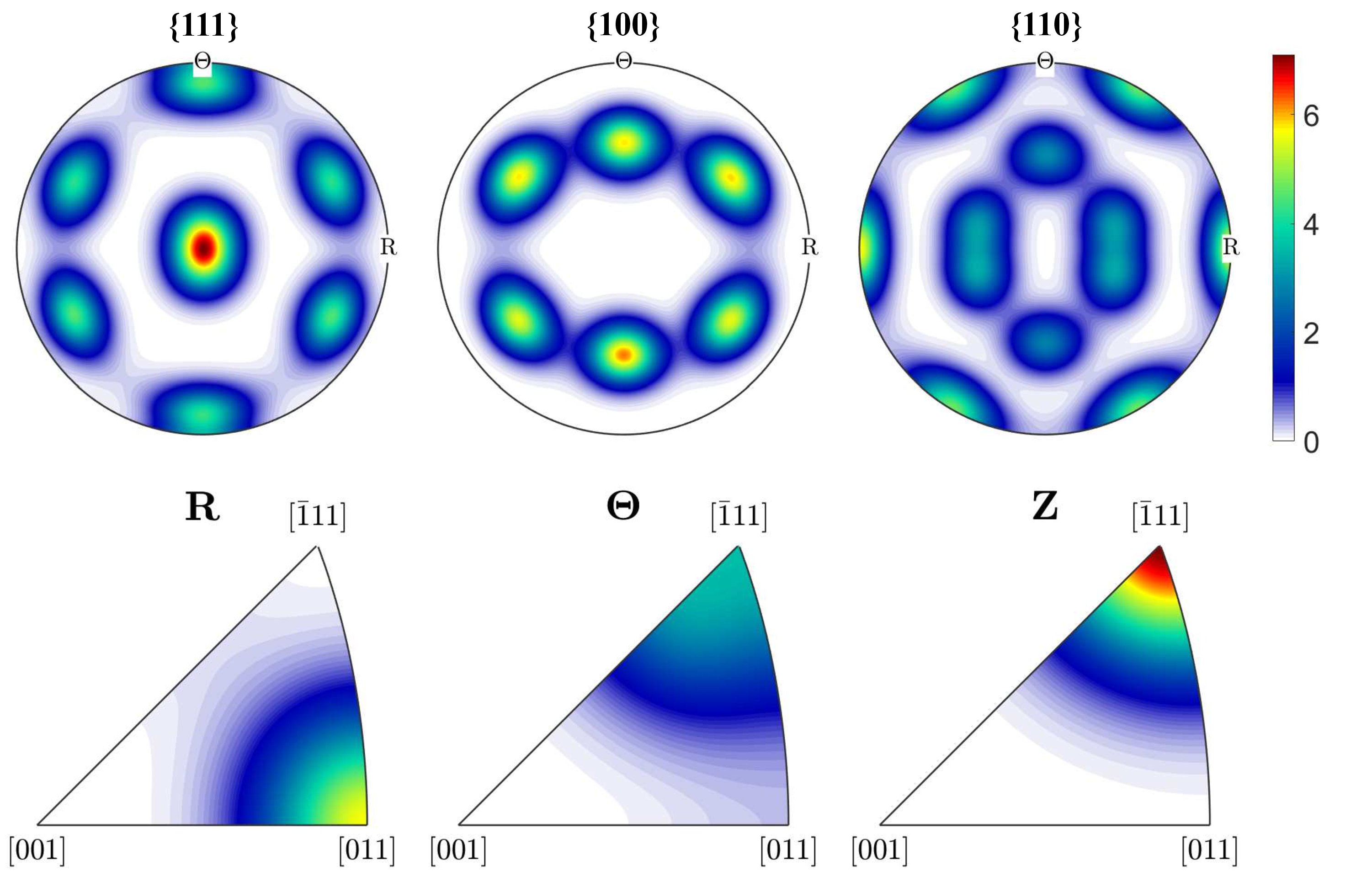}
		\label{111-Z_TEXTURE_N5000-D1_V_1000_PFs_FINAL}
	}	
	\caption{Initial $\left\langle 111\right\rangle\parallel$ Z fiber texture. Microstructure N5000-D1. Pole figures for $\left\lbrace 111\right\rbrace$, $\left\lbrace 100\right\rbrace $ and $\left\lbrace 110\right\rbrace$ plane families with respect to $(R,\Theta,Z)$ coordinate system, and inverse pole figures for R, $\Theta$ and Z directions with respect to $\left[001\right]$, $\left[011\right]$ and $\left[\bar{1}11\right]$ crystallographic axes. Results corresponding to: (a) undeformed and (b) deformed configurations, respectively. The deformed configuration is computed at the necking time. The calculations are performed for a {loading} velocity $V_{R}=1000~\mathrm{m/s}$. For interpretation of the references to color in this figure legend, the reader is referred to the web version of this article.} 
	\label{111-Z texture pole figures}
\end{figure}

Figs. \ref{111-Z_TEXTURE_RING_V_1000} and \ref{111-Z texture pole figures} include results for the initial $\left\langle 111\right\rangle\parallel$ Z fiber texture. The grains in polycrystalline materials with such texture have their crystallographic directions $\left\langle 111\right\rangle $ generally aligned with (equivalently, crystallographic planes $\left\{111\right\}$ perpendicular to) the ring axial direction Z, while the orientation of the crystallographic planes containing these directions remains random. As a result, the texture displays axial symmetry, so that the ring material is transversely isotropic, with the $R-\theta$ being the isotropy plane. As already discussed before, the $\left\langle 111\right\rangle$ fiber texture is a basic texture component found after extrusion process of initially untextured FCC materials. In such process, grains reorient to align their $\left\{111\right\}$ crystallographic planes (i.e slip planes) perpendicularly to the extrusion direction, so such texture would be observed if the ring is obtained from the extruded tube. The presentation of results follows the same scheme used for the three previous textures investigated.

Figs. \ref{111-Z_TEXTURE_RING_N5000_V_1000} and \ref{111-Z_TEXTURE_RING_N15000_V_1000} show the normalized circumferential logarithmic strain $\hat{\varepsilon}_{\theta\theta}$ versus the normalized perimeter of the ring $\hat{P}$ for calculations containing $5000$ and $15000$ grains, respectively. Decreasing the average grain size increases the number of necks and the necking time, as for the other three initial textures investigated in this paper. The number of necks for the three aggregates with $5000$ grains is $4$, $5$ and $6$ for the black, green and red curves, respectively, while $6$ excursions of strain meet the necking criterion for the three calculations with $15000$ grains. {We have checked in the field outputs of the simulation N15000-D1 that the two strain peaks in necks $4$ and $5$ of Fig. \ref{111-Z_TEXTURE_RING_N15000_V_1000} correspond to a single neck (local increase of the strain inside the neck due to grains reorientation).} Moreover, the average necking time considering the microstructures N5000-D1, N5000-D2 and N5000-D3 is $7.5\%$ smaller than for the calculations with $15000$ grains. Note also that the $\hat{\varepsilon}_{\theta\theta}-\hat{P}$ curves show a \textit{rough} profile, similarly to the calculations with initial isotropic texture and with initial $\left\langle 001\right\rangle\parallel$ R Goss texture, which indicates that the grains rotate upon deformation of the ring. The change of texture during loading is apparent in the results of the texture analysis which are shown in Figs. \ref{111-Z_TEXTURE_N5000-D1_V_1000_PFs_INITIAL} and \ref{111-Z_TEXTURE_N5000-D1_V_1000_PFs_FINAL} for the undeformed configuration and for the necking time, respectively. For instance, the spherical symmetry of the pole figures in the undeformed configuration turns into reflection symmetry, which is accompanied by a decrease in the texture intensity (the texture index ranges from $0$ to $11$ in the undeformed configuration, and from $0$ to $7.1$ at the necking time). Similarly to the previously analyzed case, the grains tend towards achieving a stable orientation, so to align one of $\left\langle 111\right\rangle$ crystal directions with the circumferential ring direction, which in this case is particularly difficult due to initial alignment of one of $\left\langle111\right\rangle$ family with the Z axis. Nevertheless, this trend is visible by a weakening of the texture intensity and spreading of the central pole at $\left\lbrace 111\right\rbrace $ pole figure. This trend is also noticeable in the inverse pole figures for the radial and circumferential directions, that were initially the same due to axial symmetry of initial texture, but have evolved differently upon loading, showing a tendency of the crystals to align the $\left\langle 011\right\rangle$ crystal direction with the radial ring direction, and the $\left\langle 111\right\rangle$  with the circumferential direction. On the other hand, the inverse pole figure for the Z direction looks qualitatively the same before and after deformation. This again illustrates texture evolution towards the stable orientation where here rotation around common direction  $\left\langle 111\right\rangle$ to align $\left\langle 011\right\rangle$ with radial direction precedes the rotation of remaining directions belonging to this family towards circumferential direction. It should be again remarked that, similarly to $\left\langle 001\right\rangle\parallel$ R Goss texture, due to ring cross-section anisotropy, the uniaxial tension of material volume of the ring in circumferential direction is not in this case equivalent to the extrusion process in this direction.

\

\begin{figure}[hbtp]
	\centering
		\includegraphics[width=19cm]{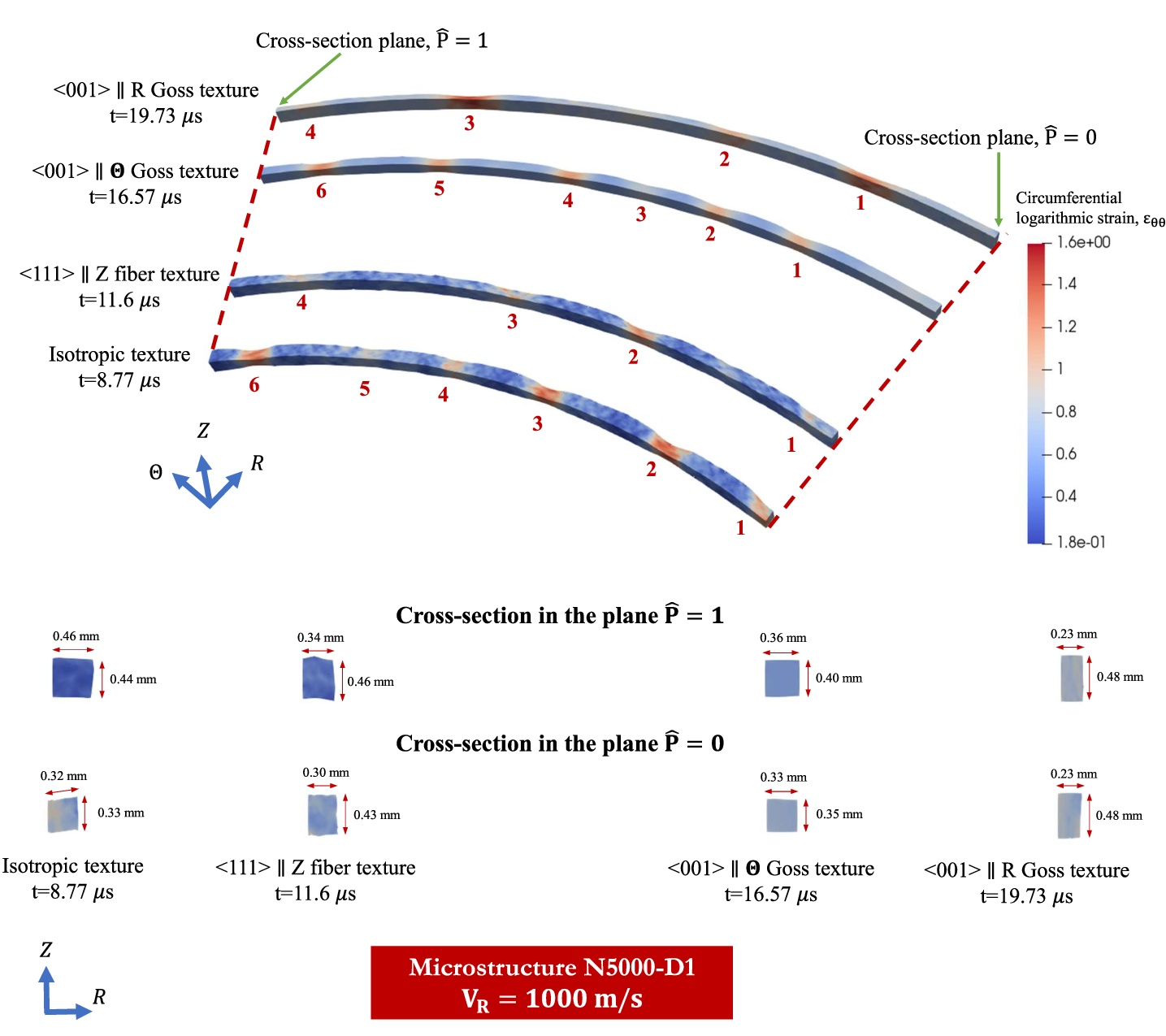}
	\caption{Contours of circumferential logarithmic strain $\varepsilon_{\theta\theta}$ for initial isotropic texture, initial $\left\langle 001\right\rangle\parallel\Theta$ Goss texture, initial $\left\langle 001\right\rangle\parallel$ R Goss texture and initial $\left\langle 111\right\rangle\parallel$ Z fiber texture. Results are shown for microstructure N5000-D1. The calculations are performed for a {loading} velocity of $V_{R}=1000~\mathrm{m/s}$. The contours correspond to the necking time.} 
	\label{CONTOURS_LEYY_1}
\end{figure}

Contours of circumferential logarithmic strain $\varepsilon_{\theta\theta}$, at the necking time, for the four initial textures investigated, are shown in Fig. \ref{CONTOURS_LEYY_1}. The results correspond to the microstructure N5000-D1. The comparison illustrates the differences in necking ductility depending on the initial texture and in the number of necks formed in the ring (the numbering of the necks is the same that in Figs. \ref{ISOTROPIC_TEXTURE_RING_N5000_V_1000}, \ref{001-THETA_RING_N5000_V_1000}, \ref{001-R_TEXTURE_RING_N5000_V_1000} and \ref{111-Z_TEXTURE_RING_N5000_V_1000}). Notice that the \textit{roughness} in the surface of the ring caused by the rotation of the grains upon deformation is especially apparent in the contours corresponding to the initial isotropic texture and the initial $\left\langle 111\right\rangle\parallel$ Z fiber texture. Moreover, the initial texture also affects the shape of the cross-section of the ring. For instance, for $\hat{P}=1$, the cross-section is \textit{roughly} square for the initial isotropic texture and the initial $\left\langle 001\right\rangle\parallel$ R Goss texture, and rectangular for the initial $\left\langle 001\right\rangle\parallel\Theta$ Goss texture and the initial $\left\langle 111\right\rangle\parallel$ Z fiber texture (elongated rectangle along Z direction, dimensions indicated in the contour plots). For $\hat{P}=0$, the shape of the ring cross-section for all anisotropic  initial textures is very similar to the results obtained for $\hat{P}=1$, while in the case of the initially isotropic material, the cross-section is sheared, most likely, due to the neck nucleated at $\hat{P}=0$. Note that the deformed shape of the ring cross-section is the manifestation of the intensity of material plastic anisotropy within this plane induced by texture. The more it departs from square shape the more intense induced anisotropy is.

\subsection{The effect of {loading} velocity}
\label{The effect of of impact velocity}

The results included in this section correspond to calculations performed with the four initial textures investigated in this paper, with microstructure N5000-D1, and for two different {loading} velocities, $V_{R}=250~\mathrm{m/s}$ and $5000~\mathrm{m/s}$ (which correspond to nominal strain rates of $1.66 \cdot 10^4~\text{s}^{-1}$ and $3.33 \cdot 10^5~\text{s}^{-1}$, respectively).

\begin{figure}[hbtp]
	\centering
	\subfloat[]{	
		\includegraphics[width=9cm]{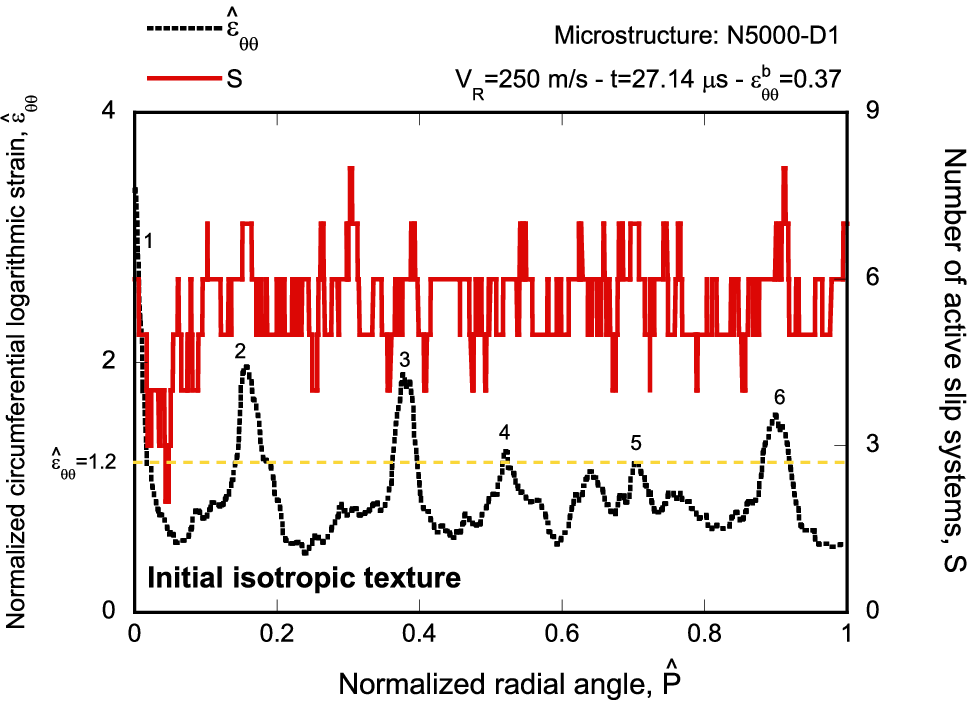}
		\label{Isotropic_Slip_250}
	}\
	\subfloat[]{	
		\includegraphics[width=9cm]{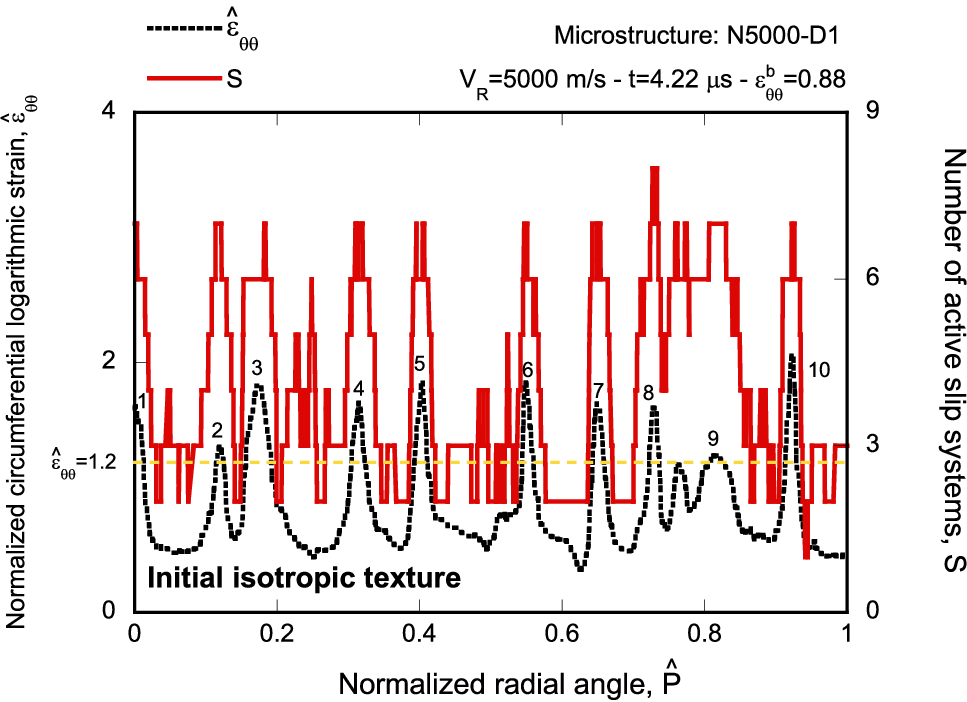}
		\label{Isotropic_Slip_5000}
	}	
	\caption{Initial isotropic texture. Normalized circumferential logarithmic strain $\hat{\varepsilon}_{\theta\theta}$ and {number of active slip systems} $S$ versus normalized perimeter of the ring $\hat{P}$ for calculations performed with microstructure N5000-D1. Results corresponding to the necking time for: (a) $V_{R}=250~\mathrm{m/s}$ and $\dot{\varepsilon}_0=1.66 \cdot 10^4~\text{s}^{-1}$, (b) $V_{R}=5000~\mathrm{m/s}$ and $\dot{\varepsilon}_0=3.33 \cdot 10^5~\text{s}^{-1}$. The necks are indicated with black numbers. The horizontal yellow dashed line corresponds to $\hat{\varepsilon}_{\theta\theta}=1.2$. For interpretation of the references to color in this figure legend, the reader is referred to the web version of this article.} 
	\label{Isotropic_Slip}
\end{figure}

Fig. \ref{Isotropic_Slip} shows the normalized circumferential logarithmic strain $\hat{\varepsilon}_{\theta\theta}$ and the {number of active slip systems $S$ (among the $12$ slip systems of the FCC crystal structure, see Table \ref{Slip_systems}) versus the normalized perimeter of the ring $\hat{P}$ for calculations performed with the initial isotropic texture. A slip system is considered to be active when $\frac{\left|{\dot{\gamma}}^\alpha\right|}{\left|\max{\left({\dot{\gamma}}^1,\ldots,{\dot{\gamma}}^{12}\right)}\right|}>0.05$.} The results correspond to the necking time. The calculation for $V_{R}=250~\mathrm{m/s}$ is shown in Fig. \ref{Isotropic_Slip_250}. The necking pattern consists of $6$ necks that are accompanied by a local increase of the slip activity. Such slip activity is a result of the developed texture in which most of the crystals orient their $\left\langle 111 \right\rangle$ directions with the circumferential ring axis, what under tension in this direction, leads to the activity of six slip systems. Moreover, minority of crystals which orient $\left\langle 001 \right\rangle$ directions with loading direction show activity of eight systems. Notice that the lowest values of slip activity appear next to the necks, so that in the transition zones between the unloaded and the strain localization sections of the ring. The number does not drop down to zero because according to the rate dependent power law (\ref{Flow_rule}) as soon as there is non-zero resolved shear stress the system will show some slip activity {(see the criterion for a slip system to be active introduced above)}. The calculation for $V_{R}=5000~\mathrm{m/s}$ is shown in Fig. \ref{Isotropic_Slip_5000}. There are $10$ necks, showing that the number of necks increases with the loading velocity. This trend is consistent with all experimental, numerical and theoretical results reported in the literature for ductile metallic rings (e.g., \cite{Altynova96}, \cite{Rusinek07} and \cite{Mercier03}), which showed that inertia effects at high strain rates favor the formation of more necks with shorter wavelength. In addition, the necking pattern is more uniform at high loading rate, the average normalized peak strain of the necks and the corresponding standard deviation for $250~\mathrm{m/s}$ and $5000~\mathrm{m/s}$ being $1.91 \pm 0.78$ and $1.68 \pm 0.23$, respectively. These results show that inertia favors that the necks grow at similar speed \citep{vaz2019comparative}. Note also that the background strain at the necking time increases with the loading rate from $0.37$ for $1.66 \cdot 10^4~\text{s}^{-1}$ to $0.88$ for $3.33 \cdot 10^5~\text{s}^{-1}$ due to the stabilizing effect of inertia which delays necking formation. Moreover, the pattern of slip activity is qualitatively very similar that for $250~\mathrm{m/s}$, with a local increase of the active slip systems at the necks, and local drops next to the necks.

\

\begin{figure}[hbtp]
	\centering
	\subfloat[]{	
		\includegraphics[width=9cm]{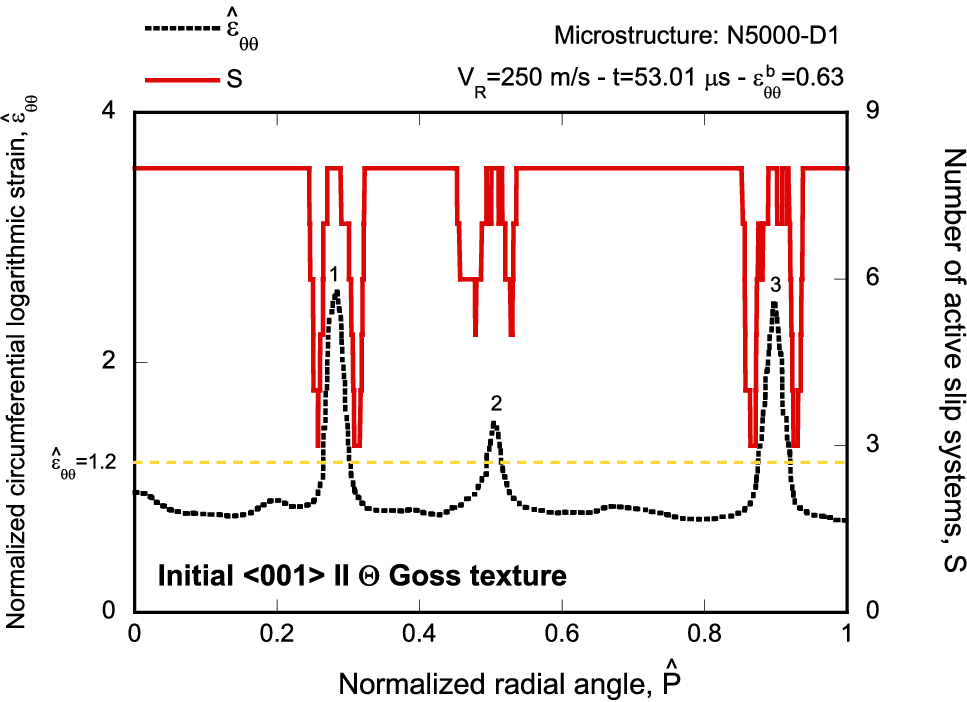}
		\label{Goss_THETA_Slip_250}
	}\
	\subfloat[]{	
		\includegraphics[width=9cm]{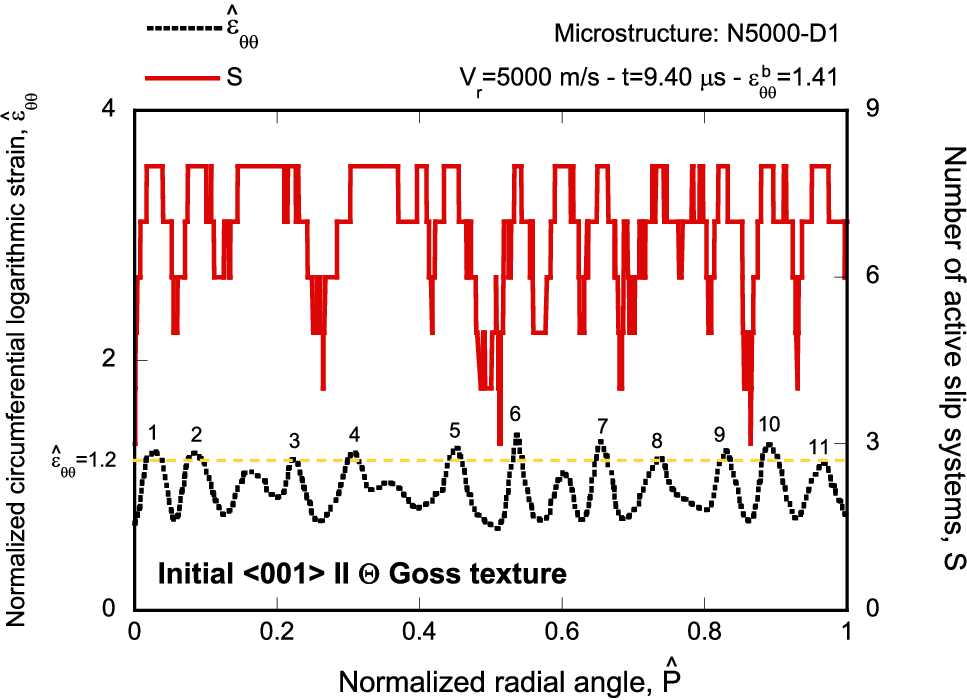}
		\label{Goss_THETA_Slip_5000}
	}	
	\caption{Initial $\left\langle 001\right\rangle\parallel\Theta$ Goss texture. Normalized circumferential logarithmic strain $\hat{\varepsilon}_{\theta\theta}$ and {number of active slip systems} $S$ versus normalized perimeter of the ring $\hat{P}$ for calculations performed with microstructure N5000-D1. Results corresponding to the necking time for: (a) $V_{R}=250~\mathrm{m/s}$ and $\dot{\varepsilon}_0=1.66 \cdot 10^4~\text{s}^{-1}$, (b) $V_{R}=5000~\mathrm{m/s}$ and $\dot{\varepsilon}_0=3.33 \cdot 10^5~\text{s}^{-1}$. The necks are indicated with black numbers. The horizontal yellow dashed line corresponds to $\hat{\varepsilon}_{\theta\theta}=1.2$. For interpretation of the references to color in this figure legend, the reader is referred to the web version of this article.} 
	\label{Goss_THETA_Slip}
\end{figure}

Fig. \ref{Goss_THETA_Slip} shows the same results for the initial $\left\langle 001\right\rangle\parallel\Theta$ Goss texture. Similar to the initial isotropic case, increasing the loading rate leads to the nucleation of more necks that grow at similar speed. The pattern of the slip activity also depends on the {loading} velocity. For $250~\mathrm{m/s}$ there are $8$ slip systems active, in agreement with the fact that the circumferential (tension) direction is aligned with the $\left\langle 001 \right\rangle$ crystal direction throughout the whole process (no lattice reorientation, see Fig. \ref{001-THETA_N5000-D1_V_1000_PFs_FINAL}), except at the sections of the ring next to the necks, in which the slip activity drops up to $3$ systems for necks $1$ and $3$. For $5000~\mathrm{m/s}$ there are more drops in the slip activity, because more necks are formed in the ring. Moreover, the background strain at the necking time increases from $0.66$ for $1.66 \cdot 10^4~\text{s}^{-1}$ to $1.41$ for $3.33 \cdot 10^5~\text{s}^{-1}$.

\

\begin{figure}[hbtp]
	\centering
	\subfloat[]{	
		\includegraphics[width=9cm]{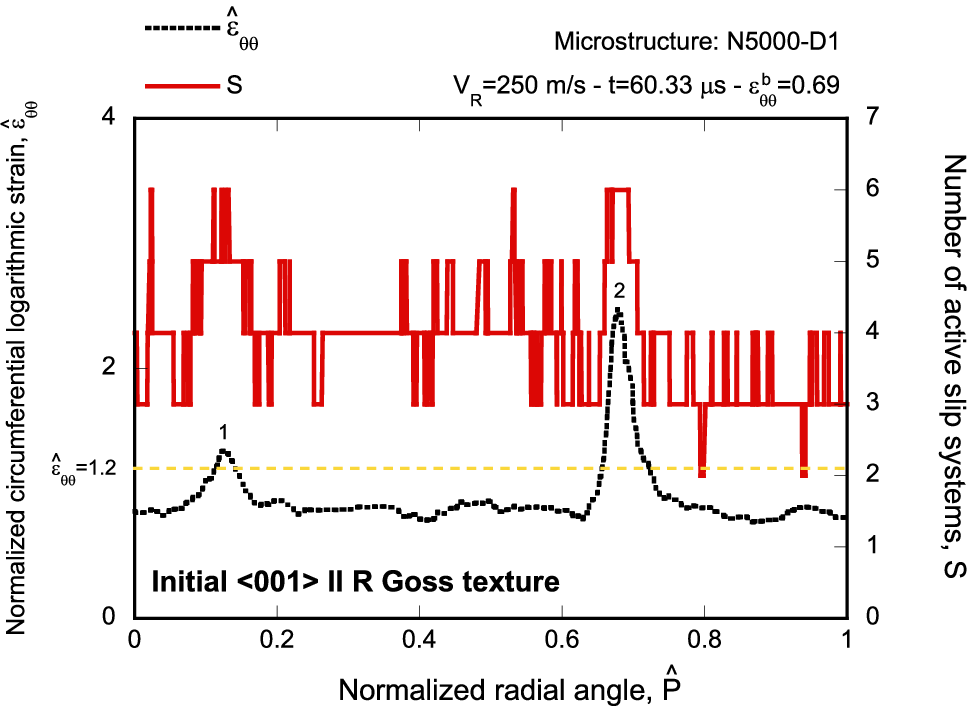}
		\label{Goss_R_Slip_250}
	}\
	\subfloat[]{	
		\includegraphics[width=9cm]{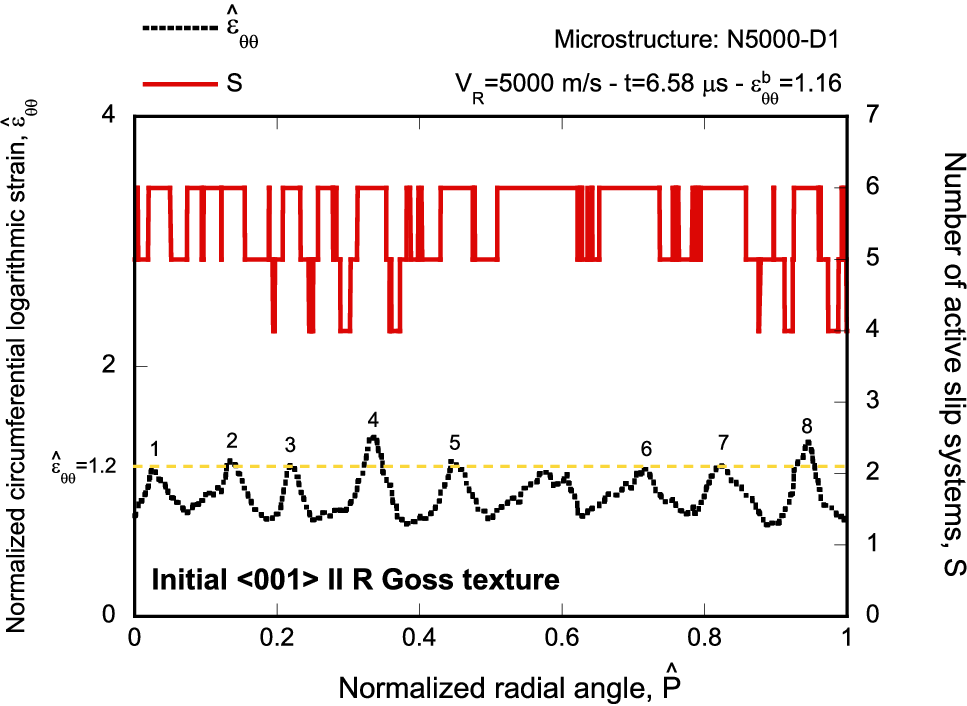}
		\label{Goss_R_Slip_5000}
	}	
	\caption{Initial $\left\langle 001\right\rangle\parallel$ R Goss texture. Normalized circumferential logarithmic strain $\hat{\varepsilon}_{\theta\theta}$ and {number of active slip systems} $S$ versus normalized perimeter of the ring $\hat{P}$ for calculations performed with microstructure N5000-D1. Results corresponding to the necking time for: (a) $V_{R}=250~\mathrm{m/s}$ and $\dot{\varepsilon}_0=1.66 \cdot 10^4~\text{s}^{-1}$, (b) $V_{R}=5000~\mathrm{m/s}$ and $\dot{\varepsilon}_0=3.33 \cdot 10^5~\text{s}^{-1}$. The necks are indicated with black numbers. The horizontal yellow dashed line corresponds to $\hat{\varepsilon}_{\theta\theta}=1.2$. For interpretation of the references to color in this figure legend, the reader is referred to the web version of this article.} 
	\label{Goss_R_Slip}
\end{figure}

Fig. \ref{Goss_R_Slip} includes the results corresponding to initial $\left\langle 001\right\rangle\parallel$ R Goss texture. We have checked that, before necking, due to the initial texture in which $\left \langle 011 \right\rangle$ crystal axes are aligned with the circumferential (tension) direction, four slip systems are active throughout the ring. Next, when approaching the necking time, due to the lattice reorientation, the number of active slip systems increases to six in the strain localization zones. In general, the patterns of slip activity for $\left\langle 001\right\rangle\parallel$ R Goss texture are qualitatively similar to the cases of initial isotropic texture and $\left\langle 001\right\rangle\parallel\Theta$ Goss texture, showing a local increase of slip activity at the necks and/or a local decrease next to the necks.  Moreover, the number of necks from $250~\mathrm{m/s}$ to $5000~\mathrm{m/s}$ increases from $2$ to $8$ (compare Figs. \ref{Goss_R_Slip_250} and \ref{Goss_R_Slip_5000}), which is slightly more than in the case of $\left\langle 001\right\rangle\parallel\Theta$ Goss texture for which the increase is $266\%$, and much more than in the case of the initial isotropic texture for which the increase is comparatively modest $\left(66\%\right)$. This result shows that the increase in the number of necks with the loading rate depends on the initial texture. Similarly, the background strain at the necking time increases with the loading rate depending on the initial texture. Namely, for the initial isotropic texture, the $\left\langle 001\right\rangle\parallel\Theta$ Goss texture and the $\left\langle 001\right\rangle\parallel$ R Goss texture, the increase in the loading velocity from $250~\mathrm{m/s}$ to $5000~\mathrm{m/s}$ leads to an increase of the background strain at the necking time of $237\%$, $223\%$ and $168\%$, respectively. 

\

\begin{figure}[hbtp]
	\centering
	\subfloat[]{	
		\includegraphics[width=9cm]{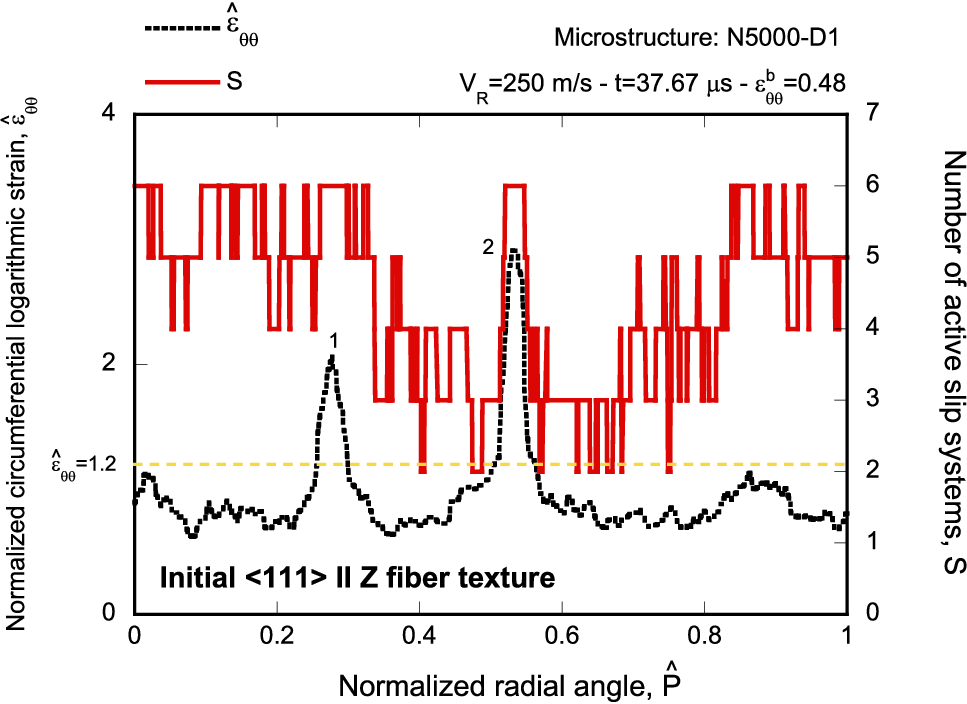}
		\label{Fiber_Z_Slip_250}
	}\
	\subfloat[]{	
		\includegraphics[width=9cm]{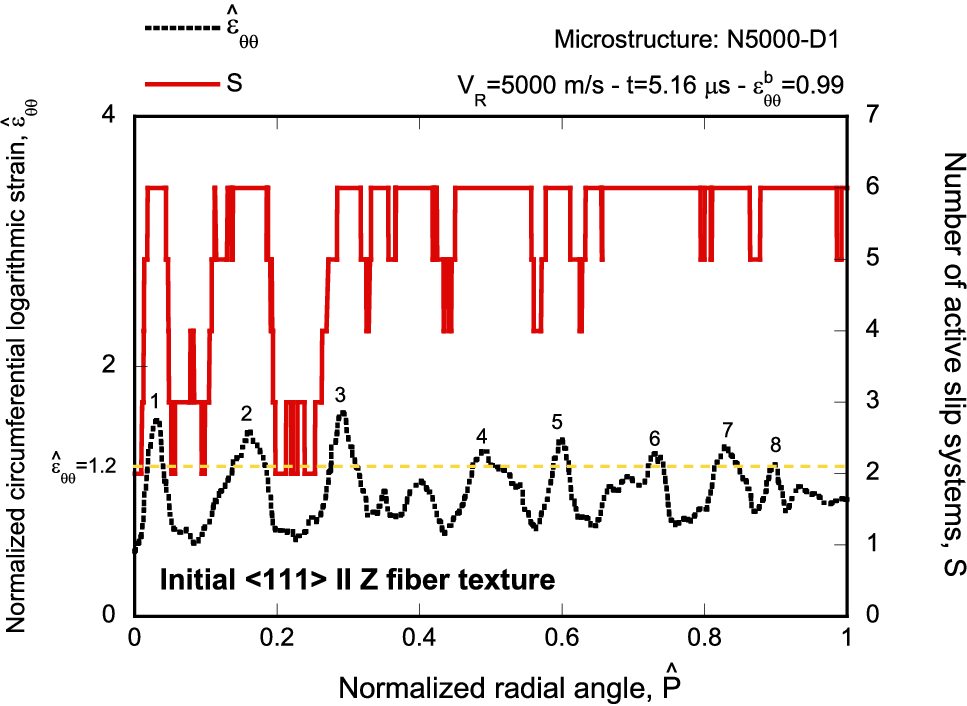}
		\label{Fiber_Z_Slip_5000}
	}	
	\caption{Initial $\left\langle 111\right\rangle\parallel$ Z fiber texture. Normalized circumferential logarithmic strain $\hat{\varepsilon}_{\theta\theta}$ and {number of active slip systems} $S$ versus normalized perimeter of the ring $\hat{P}$ for calculations performed with microstructure N5000-D1. Results corresponding to the necking time for: (a) $V_{R}=250~\mathrm{m/s}$ and $\dot{\varepsilon}_0=1.66 \cdot 10^4~\text{s}^{-1}$, (b) $V_{R}=5000~\mathrm{m/s}$ and $\dot{\varepsilon}_0=3.33 \cdot 10^5~\text{s}^{-1}$. The necks are indicated with black numbers. The horizontal yellow dashed line corresponds to $\hat{\varepsilon}_{\theta\theta}=1.2$. For interpretation of the references to color in this figure legend, the reader is referred to the web version of this article.} 
	\label{Fiber_Z_Slip}
\end{figure}

The results corresponding to initial $\left\langle 111\right\rangle\parallel$ Z fiber texture are included in Figs. \ref{Fiber_Z_Slip_250} and \ref{Fiber_Z_Slip_5000} for {loading} velocities of $250~\mathrm{m/s}$ and $5000~\mathrm{m/s}$, respectively. As in the case of initial $\left\langle 001\right\rangle\parallel$ R Goss texture, the number of necks increases from $2$ to $8$, and the slip activity increases at the necked sections, and drop next to the necks.

\section{Concluding remarks}
\label{Concluding remarks}

Finite element calculations of ductile metallic rings subjected to dynamic radial expansion have been performed for nominal strain rates varying within the range $1.66 \cdot 10^4 ~ \text{s}^{-1} \leq \dot{\varepsilon}_{0} \leq 3.33 \cdot 10^5 ~ \text{s}^{-1}$. The mechanical behavior of the material has been described with the elasto-viscoplastic single crystal constitutive model developed by \citet{marin2006}, with parameters representative of aluminium. The rings have been tessellated into an aggregate of grains using random Voronoi seeds. Two polycrystalline aggregates of $5000$ and $15000$ grains have been generated, and for each polycrystalline aggregate three different spatial distributions of grains have been considered. Each grain of the tessellation is allotted with a crystal orientation taken from one of the four different initial textures that have been investigated: isotropic texture, $\left\langle 001\right\rangle\parallel\Theta$ Goss texture, $\left\langle 001\right\rangle\parallel$ R Goss texture and $\left\langle 111\right\rangle\parallel$ Z fiber texture. The main conclusions drawn from this research are:


\begin{itemize}

\item The spatial distribution of grains affects the location of the necks.

\item The decrease of the grain size delays the formation of the necking pattern and increases the number of necks.

\item The increase of the loading rate delays necking formation, increases the number of necks and favors the formation of uniform necking patterns with necks that grow at similar speed.

\item The increase in the number of necks with the loading rate depends on the initial texture.

\item The initial texture affects the number of necks, the location of the necks, and the necking time.

\item The development of the necks is generally accompanied by a local increase of the slip activity, while the lowest number of active slip planes appears next to the necks, in the unloaded sections of the ring.

\item For $\left\langle 001\right\rangle\parallel$ R Goss texture, which has slip activity in the uniform deformation regime smaller than 5, necks formation is {generally} delayed as compared to $\left\langle 001\right\rangle\parallel \Theta$ Goss texture which shows slip activity equal to $8$ {(except for the largest loading velocity considered, probably, due to inertia effects)}.


\item The crystals rotate upon deformation for the initial isotropic texture, the $\left\langle 001\right\rangle\parallel$ R Goss texture and the $\left\langle 111\right\rangle\parallel$ Z fiber texture, showing a tendency to align the $\left\lbrace 111\right\rbrace$ orientation with the loading direction.

\item The rotation of the grains leads to \textit{rough} necking patterns due to differences in the local slip resistance of the crystals because of their different lattice orientation.

\item The initial $\left\langle 001\right\rangle\parallel\Theta$ Goss texture does not evolve during loading, leading to \textit{smooth} necking patterns because all crystals show similar slip resistance which leads to gentle transitions of strain from grain to grain.

\end{itemize}

The finite element calculations reported in this paper suggest that the initial texture can be tailored to delay dynamic necking localization, and thus to increase the formability and energy absorption capacity of ductile materials at high strain rates. {In future works, it would be interesting to consider HCP metals, in which different families of slip systems are active, which are governed by different material parameters. Hexagonal close-packed metals exhibit much stronger anisotropy induced by texture than face-centered cubic metals, and therefore it is expected that the differences in ring expansion and related strain localization observed for FCC metals will be magnified. Another direction for future works is the analysis of microstructures with smaller grain size, as well as constitutive models that include other deformation mechanisms such as twinning. In addition, it would be interesting to determine the link between the initial texture and the macroscopic response of the material to connect the necking ductility of the rings to the initial yield stress and the strain hardening of the polycrystalline aggregate (along the circumferential direction of the ring).}

\section*{Funding}

This work has received funding from the European Union's Horizon 2020 Programme (Excellent Science, Marie-Sk\l{}odowska-Curie Actions) under REA grant agreement 777896 (Project QUANTIFY). 

\

The support of National Science Centre, Poland, through the project 2021/41/B/ST8/03345 is acknowledged.

\section*{Conflict of interest} 

The authors declare that they have no known competing financial interests or personal relationships that could have appeared to influence the work reported in this paper.

\section*{Author contributions}
\label{Author contributions}

\textbf{K. E. N'souglo}: Conceptualization; Data curation; Formal analysis; Investigation; Methodology; Software; Validation; Writing - original draft; Writing - review \& editing. \textbf{K. Kowalczyk-Gajewska}: Conceptualization; Formal Analysis; ; Writing - original draft; Writing - review \& editing. \textbf{M. Marvi-Mashhadi}: Conceptualization; Methodology; Software; Writing - review \& editing. \textbf{J. A. Rodr\'{i}guez-Mart\'{i}nez}: Conceptualization; Formal analysis; Funding acquisition; Investigation; Methodology; Project administration; Resources; Supervision; Validation; Writing - original draft; Writing - review \& editing.

\appendix
	
\section{{Necking time and number of necks}}
\label{Necking time and number of necks}

{Tables \ref{Table.3}-\ref{Table.6} include the background circumferential logarithmic strain $\varepsilon^b_{\theta \theta}$ corresponding to the necking time and the number of necks $N_{neck}$ corresponding to the calculations reported in Section \ref{Results}.}

\begin{table}[hbtp]
	\begin{center}
			\begin{tabular}{c | c c  c c  c c}
					\hline
					\rowcolor{gray!25} \multicolumn{7}{c} {Initial isotropic texture} \\ 
					\hline
					\hline
					\rowcolor{gray!10} & \multicolumn{2}{c}  {$V_R=250~\text{m}/\text{s}$}  & \multicolumn{2}{c} {$V_R=1000~\text{m}/\text{s}$} & \multicolumn{2}{c} {$V_R=5000~\text{m}/\text{s}$} \\ 					
					\hline
					\hline
					& $\varepsilon^b_{\theta \theta}$ & $N_{neck}$ & $\varepsilon^b_{\theta \theta}$ & $N_{neck}$ & $\varepsilon^b_{\theta \theta}$ & $N_{neck}$ \\ 
					\hline
					 N5000-D1 & 0.37 & 6 & 0.46 & 6 & 0.88 & 10 \\
					 N5000-D2 & - & - & 0.43 & 7 & - & - \\
					 N5000-D3 & - & - & 0.46 & 7 & - & - \\
					 N15000-D1 & - & - & 0.49 & 8 & - & - \\
					 N15000-D2 & - & - & 0.44 & 7 & - & - \\
					 N15000-D3 & - & - & 0.50 & 7 & - & - \\
					\hline
				\end{tabular}
		\end{center}
	\begin{center}
			\caption{{Initial isotropic texture. Background circumferential logarithmic strain $\varepsilon^b_{\theta \theta}$ corresponding to the necking time and number of necks $N_{neck}$ for three loading velocities $V_R=250$, $1000$ and $5000~\text{m}/\text{s}$ (all the calculations reported in this manuscript)}.}
			\label{Table.3}
		\end{center}
\end{table}

\begin{table}[hbtp]
	\begin{center}
			\begin{tabular}{c | c c  c c  c c}
					\hline
					\rowcolor{gray!25} \multicolumn{7}{c} {Initial $\left\langle 001\right\rangle\parallel\Theta$ Goss texture} \\ 
					\hline
					\hline
					\rowcolor{gray!10} & \multicolumn{2}{c}  {$V_R=250~\text{m}/\text{s}$}  & \multicolumn{2}{c} {$V_R=1000~\text{m}/\text{s}$} & \multicolumn{2}{c} {$V_R=5000~\text{m}/\text{s}$} \\ 					
					\hline
					\hline
					& $\varepsilon^b_{\theta \theta}$ & $N_{neck}$ & $\varepsilon^b_{\theta \theta}$ & $N_{neck}$ & $\varepsilon^b_{\theta \theta}$ & $N_{neck}$ \\ 
					\hline
					 N5000-D1 & 0.63 & 3 & 0.74 & 6 & 1.41 & 11 \\
					 N5000-D2 & - & - & 0.75 & 5 & - & - \\
					 N5000-D3 & - & - & 0.73 & 6 & - & - \\
					 N15000-D1 & - & - & 0.78 & 8 & - & - \\
					 N15000-D2 & - & - & 0.75 & 7 & - & - \\
					 N15000-D3 & - & - & 0.76 & 7 & - & - \\
					\hline
				\end{tabular}
		\end{center}
	\begin{center}
			\caption{{Initial $\left\langle 001\right\rangle\parallel\Theta$ Goss texture. Background circumferential logarithmic strain $\varepsilon^b_{\theta \theta}$ corresponding to the necking time and number of necks $N_{neck}$ for three loading velocities $V_R=250$, $1000$ and $5000~\text{m}/\text{s}$ (all the calculations reported in this manuscript).}}
			\label{Table.4}
		\end{center}
\end{table}

\begin{table}[hbtp]
	\begin{center}
			\begin{tabular}{c | c c  c c  c c}
					\hline
					\rowcolor{gray!25} \multicolumn{7}{c} {Initial $\left\langle 001\right\rangle\parallel$ R Goss texture} \\ 
					\hline
					\hline
					\rowcolor{gray!10} & \multicolumn{2}{c}  {$V_R=250~\text{m}/\text{s}$}  & \multicolumn{2}{c} {$V_R=1000~\text{m}/\text{s}$} & \multicolumn{2}{c} {$V_R=5000~\text{m}/\text{s}$} \\ 					
					\hline
					\hline
					& $\varepsilon^b_{\theta \theta}$ & $N_{neck}$ & $\varepsilon^b_{\theta \theta}$ & $N_{neck}$ & $\varepsilon^b_{\theta \theta}$ & $N_{neck}$ \\ 
					\hline
					 N5000-D1 & 0.69 & 2 & 0.84 & 4 & 1.16 & 8 \\
					 N5000-D2 & - & - & 0.85 & 3 & - & - \\
					 N5000-D3 & - & - & 0.88 & 5 & - & - \\
					 N15000-D1 & - & - & 0.94 & 5 & - & - \\
					 N15000-D2 & - & - & 0.93 & 6 & - & - \\
					 N15000-D3 & - & - & 0.94 & 7 & - & - \\
					\hline
				\end{tabular}
		\end{center}
	\begin{center}
			\caption{{Initial $\left\langle 001\right\rangle\parallel$ R Goss texture. Background circumferential logarithmic strain $\varepsilon^b_{\theta \theta}$ corresponding to the necking time and number of necks $N_{neck}$ for three loading velocities $V_R=250$, $1000$ and $5000~\text{m}/\text{s}$ (all the calculations reported in this manuscript).}}
			\label{Table.5}
		\end{center}
\end{table}

\begin{table}[hbtp]
	\begin{center}
			\begin{tabular}{c | c c  c c  c c}
					\hline
					\rowcolor{gray!25} \multicolumn{7}{c} {Initial $\left\langle 111\right\rangle\parallel$ Z fiber texture} \\ 
					\hline
					\hline
					\rowcolor{gray!10} & \multicolumn{2}{c}  {$V_R=250~\text{m}/\text{s}$}  & \multicolumn{2}{c} {$V_R=1000~\text{m}/\text{s}$} & \multicolumn{2}{c} {$V_R=5000~\text{m}/\text{s}$} \\ 					
					\hline
					\hline
					& $\varepsilon^b_{\theta \theta}$ & $N_{neck}$ & $\varepsilon^b_{\theta \theta}$ & $N_{neck}$ & $\varepsilon^b_{\theta \theta}$ & $N_{neck}$ \\ 
					\hline
					 N5000-D1 & 0.48 & 2 & 0.57 & 4 & 0.99 & 8 \\
					 N5000-D2 & - & - & 0.59 & 5 & - & - \\
					 N5000-D3 & - & - & 0.61 & 6 & - & - \\
					 N15000-D1 & - & - & 0.59 & 6 & - & - \\
					 N15000-D2 & - & - & 0.63 & 6 & - & - \\
					 N15000-D3 & - & - & 0.65 & 6 & - & - \\
					\hline
				\end{tabular}
		\end{center}
	\begin{center}
			\caption{{Initial $\left\langle 111\right\rangle\parallel$ Z fiber texture. Background circumferential logarithmic strain $\varepsilon^b_{\theta \theta}$ corresponding to the necking time and number of necks $N_{neck}$ for three loading velocities $V_R=250$, $1000$ and $5000~\text{m}/\text{s}$ (all the calculations reported in this manuscript).}}
			\label{Table.6}
		\end{center}
\end{table}

\bibliography{Plast_instab}
\bibliographystyle{elsarticle-harv}

\end{document}